\documentclass[prb,twocolumn,aps,superscriptaddress]{revtex4-2}

\usepackage{graphicx}
\usepackage{dcolumn}
\usepackage{bm}
\usepackage{amssymb}
\usepackage{amsmath}
\usepackage{subfigure}
\usepackage{physics}
\usepackage{sublabel}
\usepackage[utf8]{inputenc}
\usepackage{hyperref}
\usepackage[usenames, dvipsnames]{color}
\usepackage[english]{babel}
\usepackage[T1]{fontenc}
\usepackage{bbm}

\hypersetup{colorlinks = true, linkcolor=blue, citecolor=blue, urlcolor=blue}

\begin{document}


\title{Momentum relaxation effects in 2D-Xene field effect device structures}
\author{Anirban Basak}
\author{Pratik Brahma}
\thanks{Current Address: Electrical Engineering and Computer Sciences, University of California, Berkeley, California 94720, USA}
\author{Bhaskaran Muralidharan}
\affiliation{Department of Electrical Engineering, Indian Institute of Technology Bombay, Powai, Mumbai-400076, India}
\email{bm@ee.iitb.ac.in}
\date{\today}
\begin{abstract}
We analyze the electric field driven topological field effect transition on 2D-xene materials with the addition of momentum relaxation effects, in order to account for dephasing processes. The topological field effect transition between the quantum spin Hall phase and the quantum valley Hall phase is analyzed in detail using the Keldysh non-equilibrium Green's function technique with the inclusion of momentum and phase relaxation, within the self-consistent Born approximation. Details of the transition with applied electric field are elucidated for the ON-OFF characteristics with emphasis on the transport properties along with the tomography of the current carrying edge states. We note that for moderate momentum relaxation, the current carrying quantum spin Hall edge states are still pristine and show moderate decay with propagation. To facilitate our analysis, we introduce two metrics in our calculations, the coherent transmission and the effective transmission. In elucidating the physics clearly, we show that the effective transmission, which is derived rigorously from the quantum mechanical current operator is indeed the right quantity to analyze topological stability against dephasing. Exploring further, we show that the insulating quantum valley Hall phase, as a result of dephasing carries band-tails which potentially activates parasitic OFF currents, thereby degrading the ON-OFF ratios. Our analysis sets the stage for realistic modeling of topological field effect devices for various applications, with the inclusion of scattering effects and analyzing their role in the optimization of the device performance. 
\end{abstract}

\maketitle


\section{\label{sec:intro}Introduction}
Electric field driven topological phase switching between a topological and a trivial phase can be harnessed in the form of topological field effect transistors (TFET) \cite{Steinberg-2010,Sireview3,Gilbert-2021}. The pursuit of such a transition in two-dimensional (2D) materials has seen significant activity \cite{Liu-2011,Chang-2012,Ionescu-2011,Fu-2014,Akhavan-2014,Sarkar-2015,Vandenberghe-2017,Simchi-2018,Xu-2019,Yang-2020,Nadeem-2021} and experimental progress \cite{Tao2015_SiFET,Fuhrer,Gilbert-2021} recently.  In 2D-Xene materials, where ``X'' refers to Si, Ge, Sb, etc., a perpendicular electric field can trigger a phase change between quantum spin Hall (QSH) phase and quantum valley Hall (QVH) phase \cite{Ezawa-2012}. The quantized conductance of the protected dissipationless current carrying edge states of the QSH phase and the insulating QVH phase can be used as ON and OFF states respectively. Previous analysis of Various 2D field effect transistor (FET) structures show promising ON-OFF currents which suggest that they can be used in high-speed and low-power applications \cite{Vandenberghe-2017,Zheng-2020,Xu-2019,Ishida-2020,Simchi-2018}. \\
\indent There has been tremendous interest in both 2D and 3D topological insulator (TI) device structures with some promising results \cite{Akhavan-2014,Chang-2012,Qian-2014,Vandenberghe-2017,Fu-2014,Hsieh-2012,Ezawa-2014,Extra-2,Extra-3,Hu-2018,Extra-6,Extra-7,Extra-9,Extra-8} in terms of the FET operation. 
In recent years, experiments have demonstrated the capacity of $Bi_2Se_3$ layers to change the conductance of its surface in response to applied electric field \cite{Steinberg-2010,Liu-2011}, although the presence of bulk states reduce the effectiveness of such device structures to be used as an FET device \cite{Steinberg-2010,Hor-2009,Cho-2011}. Few methods can be used to reduce the number of bulk states so that the surface conduction dominates \cite{Checkelsky-2011,Bansal-2012} and other methods reduce the film thickness to introduce a gap between Dirac cones to improve conductivity \cite{Taskin-2012,Thalmeier-2020}. Other device structures like gate all around or a conventional MOSFET with depletion region can also be realized with this material \cite{Zhu-2013,Satake-2018}. However, TFETs are yet to demonstrate better performance than tunnel FETs to be a viable alternative to the existing technologies \cite{Ionescu-2011,Sarkar-2015}. \\
\begin{figure}
\subfigure[]{\label{fig:schematic-a}
    \includegraphics[width=0.8\columnwidth]{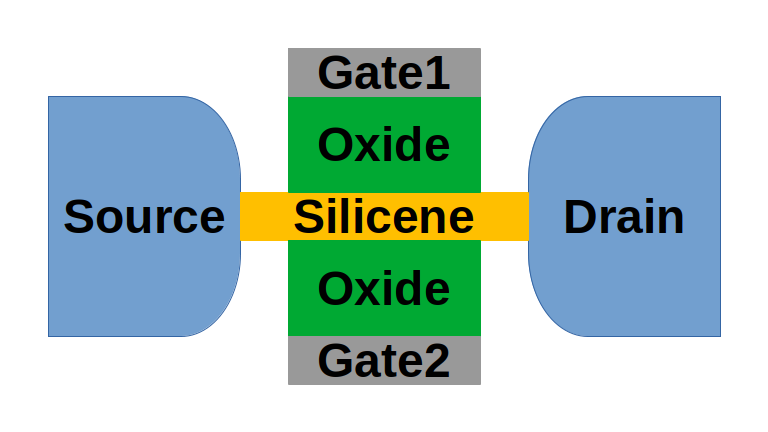}}
\hfill
\subfigure[]{\label{fig:schematic-b}
    \includegraphics[width=0.8\columnwidth]{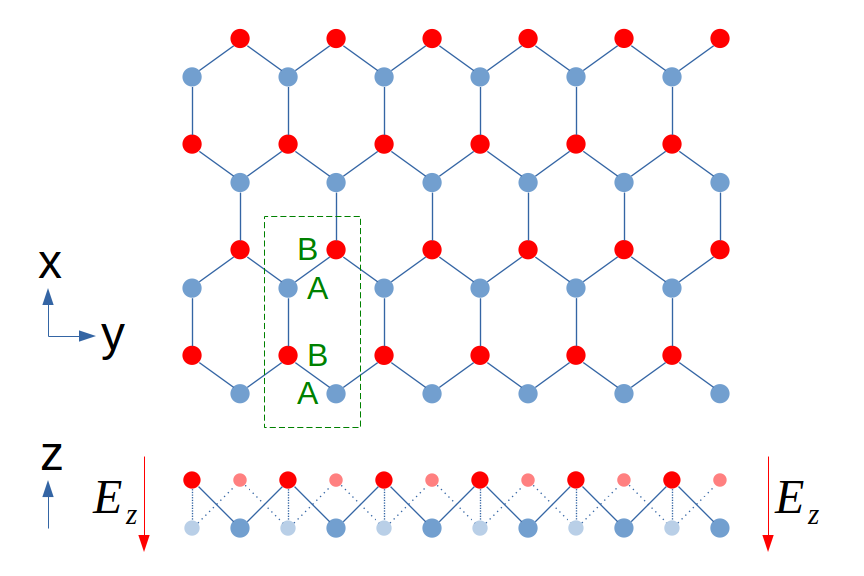}}
\hfill
\subfigure[]{\label{fig:schematic-c}
    \includegraphics[width=0.8\columnwidth]{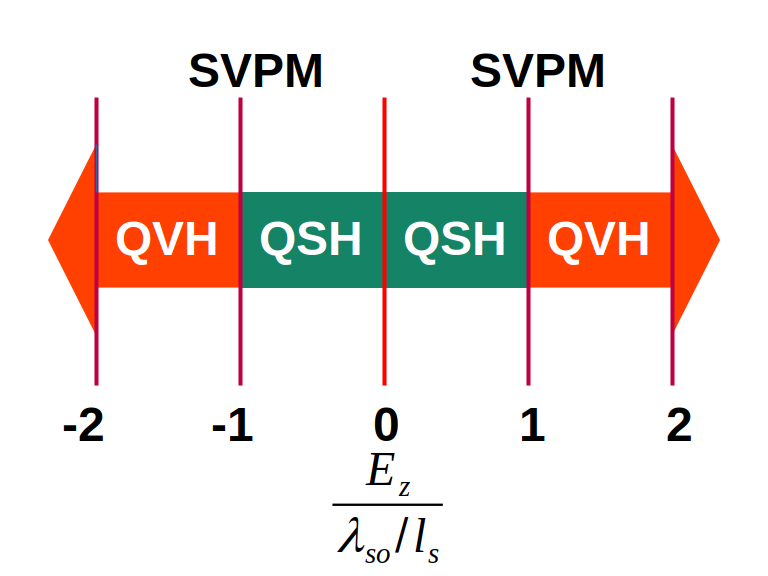}}
\caption{\label{fig:schematic}Device schematics: (a) A typical representation of the nanoribbon FET structure, with gate all around configuration. (b) Top view and side view of the buckled hexagonal lattice structure of a typical 2D-Xene channel. (c) a representative phase diagram of silicene in the presence of an applied perpendicular electric field. In (b) the green dashed rectangular box shows one unit cell and the green letters represent the sub-lattice sites marked by A and B.}
\end{figure}
\indent There is considerable interest in buckled 2D-Xene materials \cite{Sireview3,jana2021robust} with respect to engineering the QSH-QVH phase transition  \cite{Ezawa_njp_2012,Ezawa-2012,Extra-8} with the application of a perpendicular electric field. With appropriate gating structures, this field effect can be used as a topological transistor, with the switching between the ON and OFF states achieved by an applied electric field. While the QSH edge states are themselves considered dissipationless, a clear understanding of the character of this transition will necessitate the actual power considerations from a device engineering perspective. \\
\indent Most works which analyze the topological robustness of various phases focus on coherent effects, which include the addition of impurities via either random disorder potential \cite{Annica,jana2021robust,Kwant} or via self consistent fields \cite{Puddles_2016,Balram}. The consideration of finite temperature fluctuating impurities \cite{Roksana,Datta,DANIELEWICZ1984239,camsari2020nonequilibrium} and hence near elastic scattering, which are non-coherent effects, is almost absent in this context \cite{Duse_2021}. These fluctuating impurities can systematically degrade the phase coherence of electrons and give rise to dephasing, which in the parlance of quantum transport effectuates a transition from the coherent ballistic regime to the diffusive regime. The Keldysh non-equilibrium Green's function (NEGF) technique allows one to study this transition \cite{DANIELEWICZ1984239,Datta,Roksana,camsari2020nonequilibrium}, via the inclusion of such dephasing mechanisms. A very common dephasing mechanism at finite temperature is that of local fluctuating impurities that give rise to momentum relaxation, and it can be rigorously shown that increasing the magnitude of the impurity potential leads to the transition from quantized conductance to Ohm's law \cite{camsari2020nonequilibrium}. \\
\indent The object of this paper is to delve into the topological field effect transition in 2D-Xene materials with quantum transport models in the coherent ballistic regime and the non-coherent regime with dephasing effects \cite{DANIELEWICZ1984239,Datta,camsari2020nonequilibrium} resulting from momentum relaxation processes \cite{Roksana,Abhishek_TED,PhysRevApplied.8.064014,Abhishek_APL,Aniket_JAP,Aritra,Praveen,Duse_2021}. The transition between the QSH and the QVH states is analyzed in detail using the NEGF technique \cite{Datta,Meir-Wingreen-1992}. Details of the transition with applied electric field are elucidated for the ON-OFF characteristics, with emphasis on the transport properties. A deeper analysis of the stability of the current carrying QSH edge states as well as the nature of the switching process is considered in detail. We note that for moderate momentum relaxation, the edge states are still pristine and show moderate decay with propagation.\\
\indent We further elucidate the physics in the non-coherent limit by introducing two possible metrics: a) the coherent transmission and the b) effective transmission. We show that although the coherent transmission seems like an intuitive quantifying metric when transitioning from the coherent limit into the non-coherent limit, the effective transmission that is rigorously derived from the current operator paints a realistic picture when it comes to evaluating the stability of the topological edge states. We elucidate that although the coherent transmission spectrum shows a significant decrease in transmission under elastic dephasing, the effective transmission in such cases remain almost close to unity, preserving the QSH property and demonstrating that the device can withstand a lot of edge defects and still maintain its ON-current. Further, we also point out the important aspect of band-tail formation \cite{Overhauser,Tillmann}, which can lead to band-gap narrowing effects in the QVH phase, a remarkable effect that can arise even in the absence of in-elastic scattering effects, which can lead to a degradation in the performance by enhancing the OFF current. The methods detailed in this paper can also be extended to include various effects of quantum dot formation and the related electron-electron interactions \cite{Puddles_2016,Muralidharan_2008}, which typically case phase randomization without momentum relaxation \cite{Roksana}, leading toward realistic modeling of topological field effect devices for various applications. \\
\indent The organization of this article is as follows: In Sec.~\ref{sec:setup-formalism}, we describe the the structure of our device, the formalism and the transport calculations using the Keldysh NEGF technique. In Sec.~\ref{sec:results} we discuss the bandstructure, local density of states, transmission spectra, and the transport properties of our device. In Sec.~\ref{sec:conclusion} we summarize our work.

\section{\label{sec:setup-formalism}Device set-up and formalism}

\subsection{\label{sec:hamiltonian}Hamiltonian}
The device structure can be visualized as a single layer of silicene placed on top of an oxide as schematized in Fig.~\ref{fig:schematic-a}. The buckled hexagonal lattice as shown in Fig.~\ref{fig:schematic-b} is a quasi-2D structure, where if seen from top the atoms seems to be arranged in a hexagonal manner on a 2D plane. However, due to the nature of the bonds and the bond angle, each atom is deviated from this common plane by a distance of $+l_s$ or $-l_s$, i.e., the two planes of atoms are separated by a total distance of $2l_s$. To maintain maximum structural stability, the nearest neighbours are displaced in opposite directions. There are four atoms in one unit cell of hexagonal lattice, arranged in lattice sites in ABAB fashion, starting from bottom-center (A) towards top-center (B) traversing clockwise through two more atoms (B and A).

We start with a generalized Hamiltonian of buckled hexagonal lattice in tight-binding representation \cite{Ezawa-2012,Ezawa-2014},
\begin{widetext}
\begin{eqnarray}
\label{eqn:hamiltonian}
H=
-t
\sum\limits_{\langle i,j\rangle \sigma}
c_{i\sigma}^{\dagger}c_{j\sigma}
+i\frac{\lambda_{SO}}{3\sqrt{3}}
\sum\limits_{\langle\langle i,j\rangle\rangle\sigma\sigma'}
\nu_{ij}c_{i\sigma}^{\dagger}\sigma_{\sigma\sigma'}^{z}c_{j\sigma'}
+i\lambda_{R1}(E_z)
\sum\limits_{\langle i,j\rangle\sigma\sigma'}
c_{i\sigma}^{\dagger}(\bm{\sigma} \times \mathbf{\hat{d}_{ij}})_{\sigma\sigma'}^{z}c_{j\sigma'} \nonumber\\
-i\frac{2}{3}\lambda_{R2}
\sum\limits_{\langle\langle i,j\rangle\rangle\sigma\sigma'}
\mu_i c_{i\sigma}^{\dagger}(\bm{\sigma} \times \mathbf{\hat{d}_{ij}})_{\sigma\sigma'}^{z}c_{j\sigma'}
-l_s E_z
\sum\limits_{i\sigma}
\mu_i c_{i\sigma}^{\dagger}c_{i\sigma},
\end{eqnarray}
\end{widetext}
where $c_{i\sigma}^{(\dagger)}$ represents annihilation (creation) operator of a spin $(\sigma =\uparrow / \downarrow)$ spin at the site $i$, $\lambda_{SO}$, $\lambda_{R1}(E_z)$ and $\lambda_{R2}$ represent the intrinsic spin-orbit coupling, the first Rashba spin-orbit coupling and the second Rashba spin-orbit coupling respectively. For the intrinsic spin-orbit coupling term, $\nu_{ij} = +1$, if the next nearest neighbour hopping is anti-clockwise and $\nu_{ij} = -1$, if the next nearest neighbour hopping is clockwise. And for the perpendicular electric field term and the second Rashba spin-orbit coupling term, $\mu_i = +1$ for lattice site A and $\mu_i = -1$ for lattice site B. The vector $\mathbf{d_{ij}}$ is the distance between the site $i$ and $j$. The $l_s$ and $E_z$ are the buckle height and perpendicular electric field respectively \cite{Ezawa-2012}.\\
\indent The first term in the Hamiltonian that of an ordinary hexagonal lattice like graphene and gives rise to the Dirac cones at the well known $K$ and $K'$ points. The intrinsic spin-orbit coupling term, $\lambda_{SO}$ , opens the gap between the Dirac cones at $K$ and $K'$ points \cite{Konschuh-2010}. The third term is the first Rashba spin-orbit coupling, $\lambda_{R1}(E_z)$, which is quite negligible for low electric field near the Dirac points.  The second Rashba spin-orbit coupling term, $\lambda_{R2}$, although not zero, is quite small. The presence of electric field removes the valley degeneracy between site A and B, and thus with the application of electric field we see a separation of spins into two distinct valleys. For silicene, the tight-binding hopping parameter $t=1.6 eV$. In our case, we take a large $\lambda_{SO} = 0.32eV$, to have a pronounced energy gap to work with. We also take $\lambda_{R1}(E_z) = 0eV$ and $\lambda_{R2} = 0.06eV$. The buckling height is set at $l_s = 0.23\mathring{A}$.\\
\indent Our work is centered around the topological transition between the QSH state and the QVH state, with the application of the electric field $E_z$ as schematized in Fig. ~\ref{fig:schematic-c}. The spin valley polarized metal (SVPM) state occurs in the intermediate state between the two phases in such a topological field switching. In order to analyze the nature of this switching process, we utilize the NEGF formalism, which we now detail below. 
\subsection{\label{sec:negf}The Keldysh NEGF formalism}
The analysis of transport for our set up is performed via the Keldysh NEGF formalism and the Landauer-B\"uttiker formalism. 
The Hamiltonian matrix, $[H]$, is constructed with the tight-binding matrix elements, $[\alpha]$ and $[\beta]$, where $[\alpha]$ represents the Hamiltonian of a single column of unit cells lying along the transverse direction of the device and $[\beta]$ represents the overlap matrix between two such columns \cite{Abhishek_APL,Abhishek_TED}. In a device connected with two leads, the starting point of the calculation is the energy resolved retarded Green's function of the device region in its matrix representation given by: 
\begin{equation}
\label{eqn:Greens-function}
[G^r(E)] = [(E+i\eta)I-H-U-\Sigma_L-\Sigma_R-\Sigma_s]^{-1}, \\
\end{equation}
where, $[I]$ is an identity matrix of the dimensions of the device Hamiltonian, $\eta$ is a small damping parameter, $U$ is the random impurity potential, $[\Sigma_{L(R)}]$ and $[\Sigma_s]$ are respectively the self-energy matrices of the contacts $L/R$ and the scattering self energy matrices representing the momentum dephasing processes respectively. 
The contact self-energies are calculated using the recursive Green's function technique where any geometry can be accommodated \cite{Datta} and the scattering self energy is evaluated using the self-consistent Born approach to be described below. In the definitions to follow, the energy argument is assumed implicit.
\begin{figure*}
\subfigure[]{\label{fig:bands-qsh}
    \includegraphics[width=0.3\textwidth]{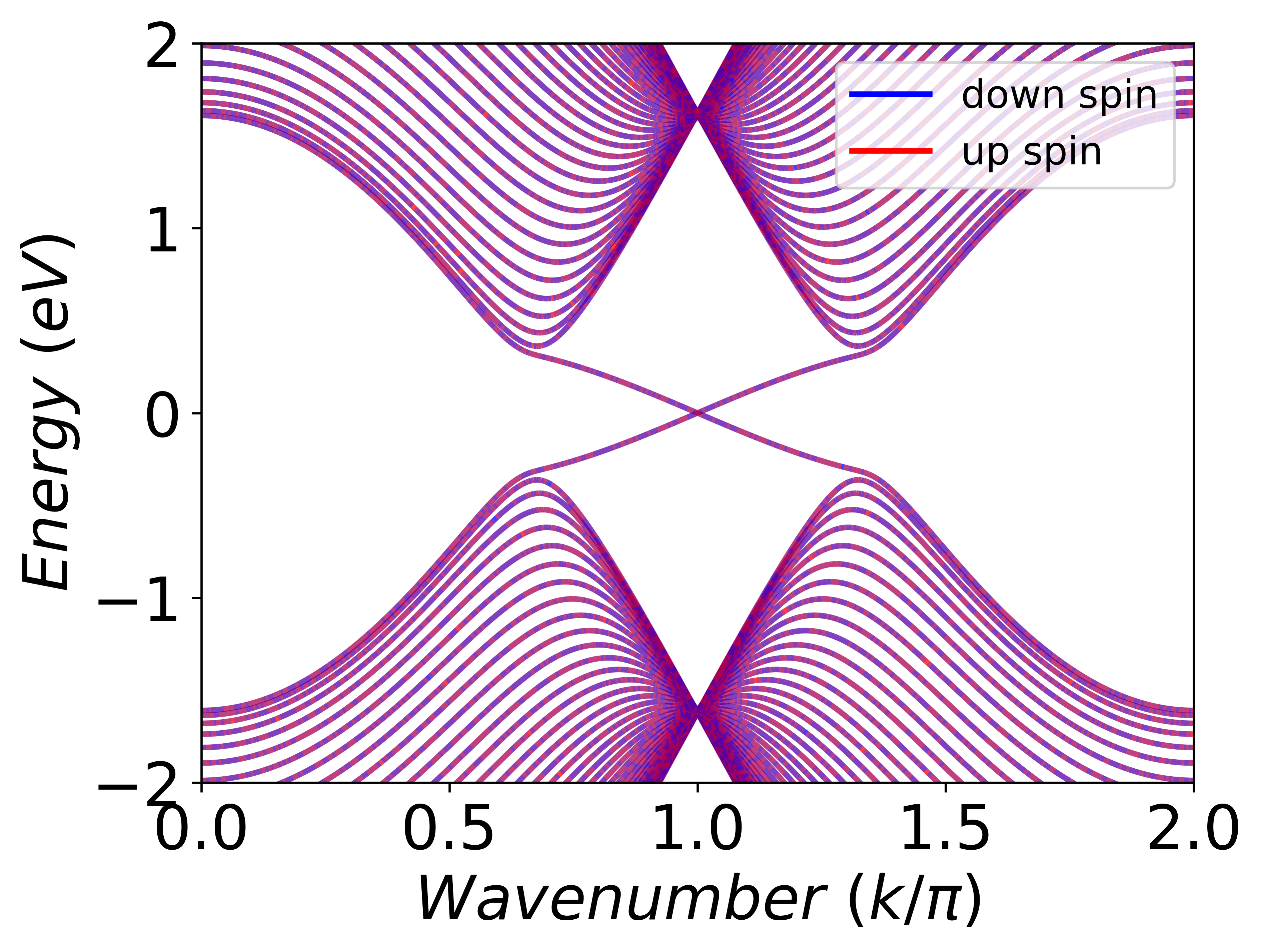}}
\subfigure[]{\label{fig:bands-svpm}
    \includegraphics[width=0.3\textwidth]{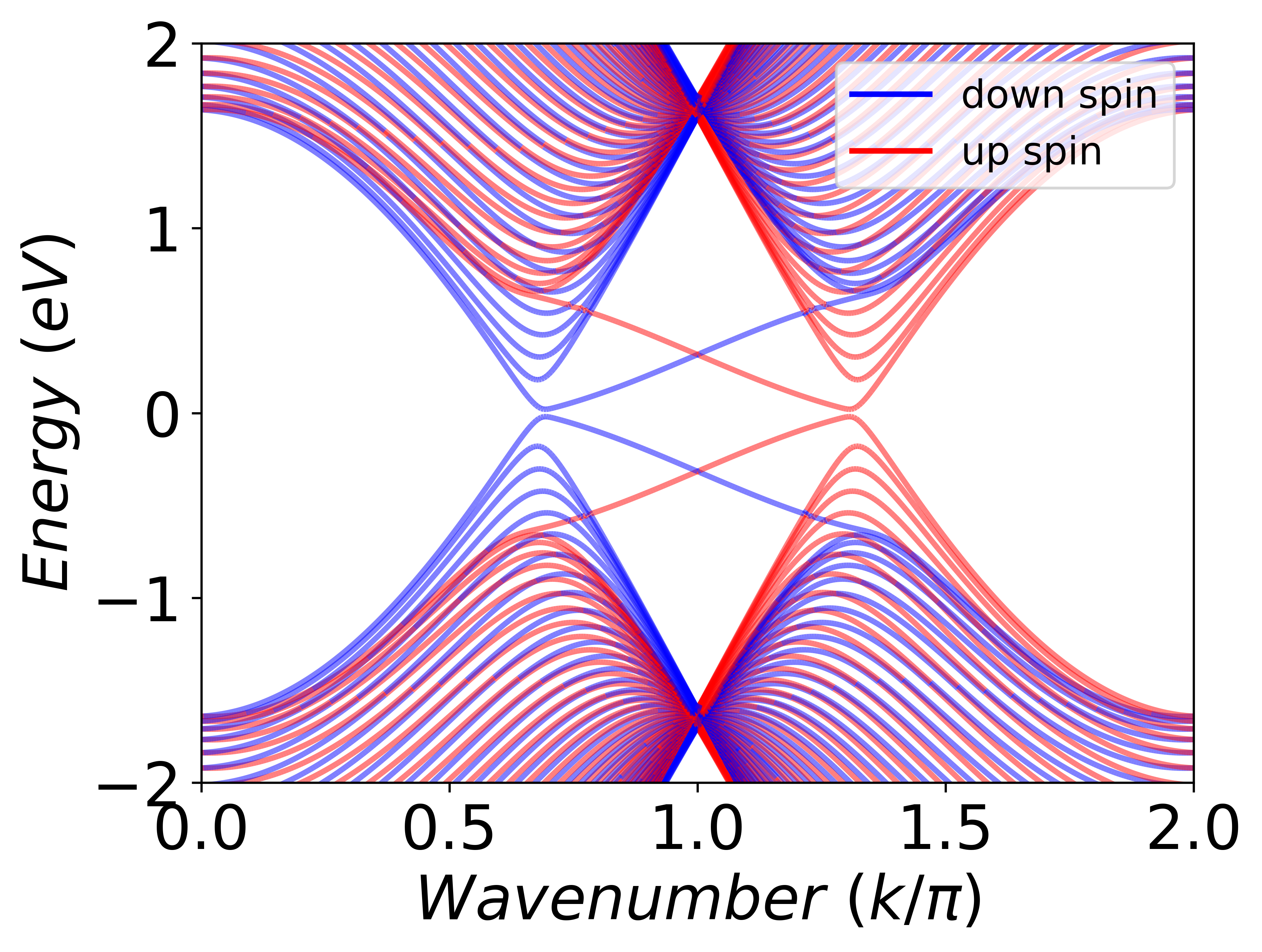}}
\subfigure[]{\label{fig:bands-qvh}
    \includegraphics[width=0.3\textwidth]{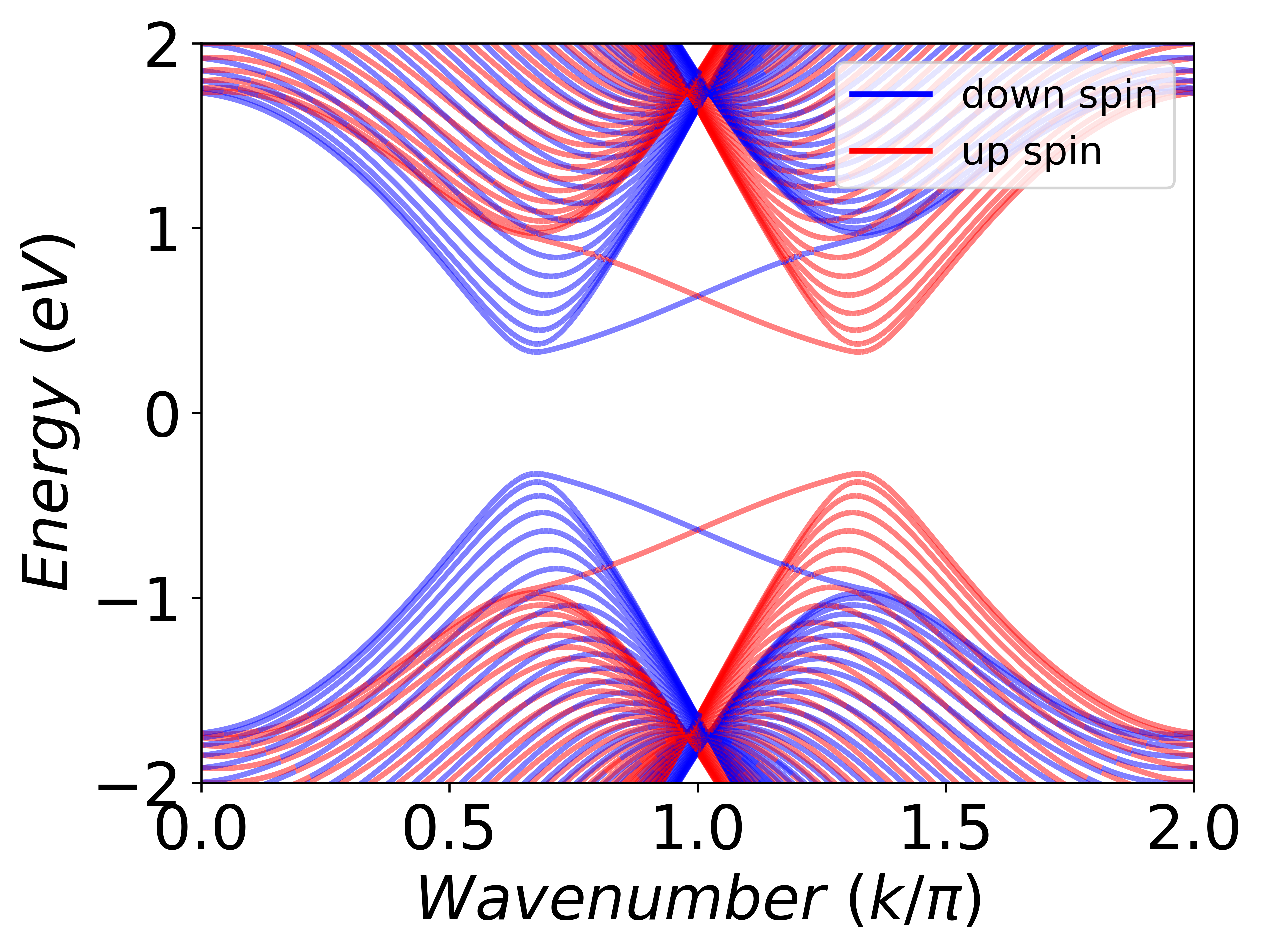}}
\caption{\label{fig:bands} Bandstructures of (a) the QSH, (b) the SVPM and (c) the QVH phases of a pristine silicene nanoribbon. The QSH phase shows the existence of a spin-degenerate edge state. The SVPM and QVH phase shows the up-spin and down-spin bands separating into two different valleys under the application of the external electric field. No edge states can be found around the Fermi energy in the case of the QVH phase. The SVPM phase represents a boundary phase between the conducting QSH phase and the insulating QVH phase.}
\end{figure*}
Using the self energies, the broadening matrix, $\Gamma$, of the contacts $L,R$ and the scattering processes are defined as,
\begin{eqnarray}
\label{eqn:lead-broadening}
[\Gamma_{L,R}] &= i[\Sigma_{L,R} - \Sigma_{L,R}^{\dagger}] \\
\label{eqn:scatterer-broadening}
[\Gamma_s] &= i[\Sigma_s - \Sigma_s^{\dagger}].
\end{eqnarray}
The spectral function, $A$, is obtained from \eqref{eqn:Greens-function} as,
\begin{eqnarray}
\label{eqn:spectral-function}
[A(E)] = i[G^r - G^a] = [G^r][\Gamma_L + \Gamma_R + \Gamma_s][G^a].
\end{eqnarray}
The diagonal elements of the spectral function are related to the local density of states (LDOS). We also require the correlator, which is related to the lesser Green's function given by
\begin{equation}
\label{eqn:electron-correlation}
[G^{<}] = [G^r][\Sigma^{<}_L + \Sigma^{<}_R + \Sigma^{<}_s][[G^a],
\end{equation}
where $\Sigma^{<}_{L(R)} = -i \Gamma_{L(R)} f_{L(R)}$ represent the lesser (in-scattering) self energy of the left (right) contact, with $f_{L(R}= f(E-\mu_{L(R)})$ is the Fermi-Dirac distribution in the left (right) contact described via their respective electrochemical potentials $\mu_{L(R)}$. The quantity $\Sigma^{<}_s$ that represents the lesser self energy of the scattering process is evaluated self-consistently, as will be detailed below. 
The retarded self energy and the lesser self energy for the momentum dephasing process is given below from \eqref{eqn:Greens-function} and \eqref{eqn:electron-correlation},
\begin{eqnarray}
\label{eqn:dephasing-self-energy}
\Sigma^r_s(i,j) =& \bar{D}(i,j)G^r(i,j) \\
\label{eqn:dephasing-electron-inscattering}
\Sigma^{<}_s(i,j) =& -i \bar{D}(i,j)G^<(i,j),
\end{eqnarray}
from which, we can define the in-scattering function  $[\Sigma^{in}_s] = -i [\Sigma^{<}_s]$ and the electron correlation function $[G^n] = -i [G^{<}]$.
These quantities are determined by the dephasing matrix, $\bar{D}(i,j)$, which is given by
\begin{eqnarray}
\label{eqn:momentum-dephasing=1}
\bar{D}(i,j) =& \braket{U_s(i)}{U^*_s(j)}.
\end{eqnarray}
We define the dephasing matrix that determines the correlation between the random scattering potentials $U_s$, distributed along the sites of the device region is defined as
\begin{eqnarray}
\label{eqn:momentum-dephasing=2}
\bar{D}(i,j) =& d_m \delta_{ij},
\end{eqnarray}
where $d_m$ is the strength of momentum relaxation, which will be added by hand to quantify the extent of dephasing. This is a homogeneous model with uniformly distributed, elastic, and spatially uncorrelated interactions, resulting in a diagonal form of $\bar{D}(i,j) = d_m \delta_{ij}$. This model discards the off-diagonal elements of the Green’s
function, thus relaxing both the phase and momentum of quasiparticles in the nanowire. \cite{DANIELEWICZ1984239,Datta}. The quantity $d_m$ typically represents the magnitude squared of the
fluctuating scattering potentials. In this work, the parameter $d_m$ is modulated so that by gradually increasing it, one can transition from the coherent ballistic limit to the diffusive limit. This model can also be extended to include non-local fluctuations via the spatial correlations of the impurity potentials \cite{DANIELEWICZ1984239,Datta}.\\
\indent Our solution for the self-energies for the dephasing process is based on the self-consistent Born approximation \cite{Datta}, in which we solve \eqref{eqn:dephasing-self-energy} and \eqref{eqn:dephasing-electron-inscattering} self consistently with \eqref{eqn:Greens-function} and \eqref{eqn:electron-correlation} respectively to obtain the full solution of the retarded and the lesser Green's function following the desired numerical convergence \cite{camsari2020nonequilibrium}. Following this, we obtain the current operator \cite{Datta} as
\begin{equation}
I^{op}_{L}= {\bf{Re}} \left({\bf{Tr}}\left[([\Sigma^{in}_L] [A]) - ([\Gamma_L] [G^n])\right] \right).
\label{curr_operator}
\end{equation}
\indent In order to quantify the effect of dephasing, inspired by the Landauer-B\"uttiker form, it is quite tempting to define a {\it{coherent transmission}}, $T_{coh}$, as a measure of electronic channels through which electrons can travel ballistically from left lead to right lead without encountering any scattering in the channel. It is obtained from \eqref{eqn:lead-broadening} and \eqref{eqn:Greens-function} as,
\begin{eqnarray}
\label{eqn:coherent-transmission}
T_{coh} = {\bf{Re}}({\bf{Tr}}(\Gamma_L G^r\Gamma_R G^a)),
\end{eqnarray}
where ${\bf{Tr}}$ stands for the matrix trace operation, and ${\bf{Re}}$ stands for the real part of the complex number obtained. However, using the current operator, we can define yet another transmission, which we term as the effective transmission, $T_{eff}$ as 
\begin{eqnarray}
\label{eqn:effective-transmission}
T_{eff} = \frac{1}{f_L - f_R} {\bf{Re}} \left({\bf{Tr}}\left[([\Sigma^{in}_L] [A]) - ([\Gamma_L] [G^n])\right] \right),
\end{eqnarray}
that represents the net number of channel for electrons to travel from left lead to right lead after encountering any number of scattering in the channel. We must note that in the presence of dephasing, the current operator, as noted in \eqref{curr_operator} cannot be written in the Landauer-B\"uttiker form \cite{Duse_2021,camsari2020nonequilibrium}. Due to this, we will establish in the upcoming section that the latter definition makes better sense as a transmission metric to analyze channels under dephasing. 
\section{\label{sec:results}Results}
\subsection{\label{sec:bands}Bandstructure}
Let us first consider the bandstructures of the three phases involved, namely, the QSH phase, the SVPM phase and the QVH phase, as pointed out in the schematic of the phase transition in Fig.~\ref{fig:schematic-c}, and qualitatively understand the role of the perpendicular electric field $E_z$ in effectuating the transitions.
The hexagonal arrangements of atoms produces Dirac cones at the six high symmetry points in reciprocal space in the Brillouin zone. Since we have a zigzag termination of atoms along the longitudinal direction, only two $K$, $K'$ symmetry points fall within the first Brillouin Zone. In the absence of spin-orbit coupling, the two Dirac cones touch each other at $K$ and $K'$ symmetry points.\\
\indent The intrinsic spin-orbit coupling causes the bulk states of the two Dirac cones at the $K$ and $K'$ symmetry points to move away from the $\Gamma$ point by an equal amount and also open a gap. The energy gap induced depends on the strength of the intrinsic spin-orbit coupling. The effect of the second Rashba spin-orbit coupling is a relationship between the transverse spin and crystal momentum, which manifests as change in band-structure depending on the crystal momentum. This change is different for different crystal momentum, i.e., different band number or longitudinal wavenumber. This change is seen more prominently in the QVH phase when the spin-up or down edge states feature a band curvature which makes their bands move towards or away from the bulk conduction or valence band edges in the vicinity of the $K$ and  $K'$ points.\\
\indent Since the orbital momentum in 2D hexagonal lattices can have two opposite orientations, the electrons at $K$ and $K'$ symmetry points can experience two equal but opposite magnetic moments. When the electron spin is coupled to this orbital momentum through spin-orbit coupling, the spin-valley degeneracy between the pairs ($K'\uparrow$,$K\downarrow$) and ($K'\downarrow$,$K\uparrow$) is lifted and a gap appears between their energy bands at $K$ and $K'$ points. Due to the nature of this effect, the electrons with the same spin but moving clockwise or counterclockwise within the hexagonal lattice will experience opposite magnetic moments.\\
\indent Our simulations on the bandstructure of the three phases are based on a silicene nanoribbon structure twenty unit cells wide and are summarized in Fig.~\ref{fig:bands}. The red and blue bands represent energy bands for the up-spin and the down-spin electrons respectively. We can see that for $|E_z| < \lambda_{SO}/l_s$, we have the QSH phase, where we have two edge states with spin degeneracy but of opposite directions as shown in Fig.~\ref{fig:bands-qsh}. The separation between the upper (conduction) and lower (valance) bulk bands in our simulation is about $0.723eV$ which is just a little over $2\lambda_{SO}=0.64eV$.\\
\indent With the application of a perpendicular electric field, electrons on all next nearest neighbouring sites gain an equal amount of energy, whereas those on the nearest neighbouring sites gain an opposite amount. Since, spin-orbit coupling also only works on the next nearest neighbouring sites, these two terms have a combined effect on the bands, shifting them vertically up or down, depending on the spin ($s$) and valley ($k$) of the electrons. Without non-zero spin-orbit coupling, the effect of the perpendicular electric field will not result in the separation of the bands.\\
\indent For small $E_z$ ($|E_z| < \lambda_{SO}/l_s$), the bands of up or down-spin will either come close together or move away from the Fermi energy. This causes the bands to separate into up-spin and down-spin bands. At the same time, for the edge states, the point at which the two opposite spin bands cross each other at the Fermi level separates in wavenumber, giving rise to two bands crossing the Fermi level at different $k$-points. This continues until $|E_zl_s| = \lambda_{SO}$, when, for a particular spin, the preexisting bulk gap from the spin-orbit effect vanishes due to the effect of the electric field at either the $K$ or the $K'$ point and the two crossings of the two opposite spin edge states merge at either the $K$ or the $K'$ point at the Fermi energy. Thus, for $|E_z| = \lambda_{SO}/l_s$, we have the SVPM phase, where the up and down-spin bands are separated into two valleys of opposite wavenumber as shown in Fig.~\ref{fig:bands-svpm}. The SVPM is a metallic state where the Fermi level lies touching both conduction and valance bands and serves as the intermediate phase where the gap begins to open.\\
\indent For large $E_z$ ($|E_z| > \lambda_{SO}/l_s$), the overall effect of the electric field dominates and it pushes both up and down-spin bands away from the Fermi energy. The edge spin bands also move away from each other along with the bulk bands and a band-gap opens. Thus, at higher electric fields $|E_z| > \lambda_{SO}/l_s$, we have the QVH phase as shown in Fig.~\ref{fig:bands-qvh}. The up and down spin bands are still separated, but now there is an energy gap between the upper (conduction) bands and lower (valence) bands. The Hall conductance in such case will only depend on the electrons at one of the two valleys. A phase diagram is shown in Fig.~\ref{fig:schematic-c} to depict the various phases of the material with respect to the applied perpendicular electric field ($E_z$). 
\subsection{\label{sec:ldos}Coherent Transport}
\begin{figure}
\subfigure[]{\label{fig:ldos-qsh}
    \includegraphics[width=0.7\columnwidth]{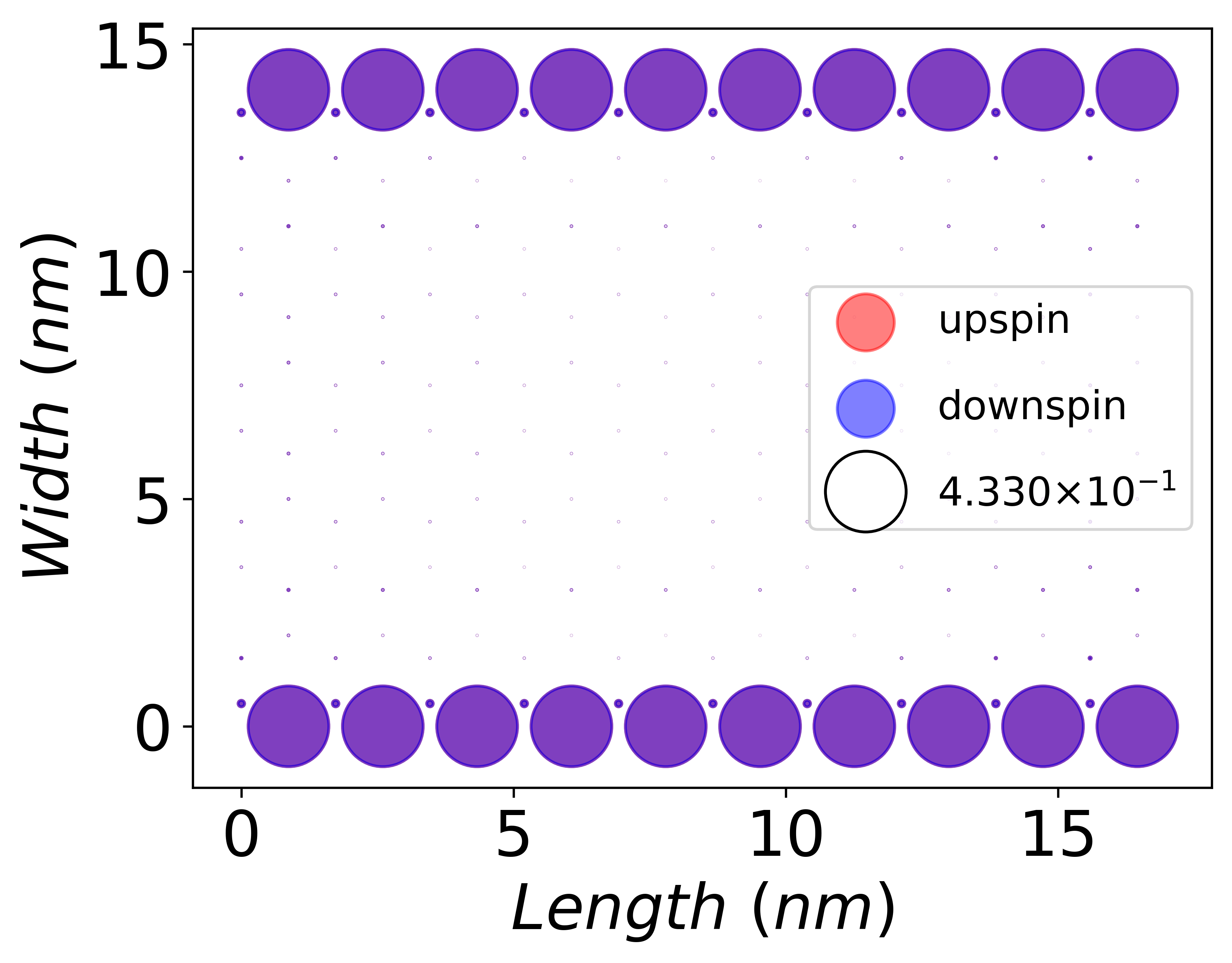}}
\hfill
\subfigure[]{\label{fig:ldos-svpm}
    \includegraphics[width=0.7\columnwidth]{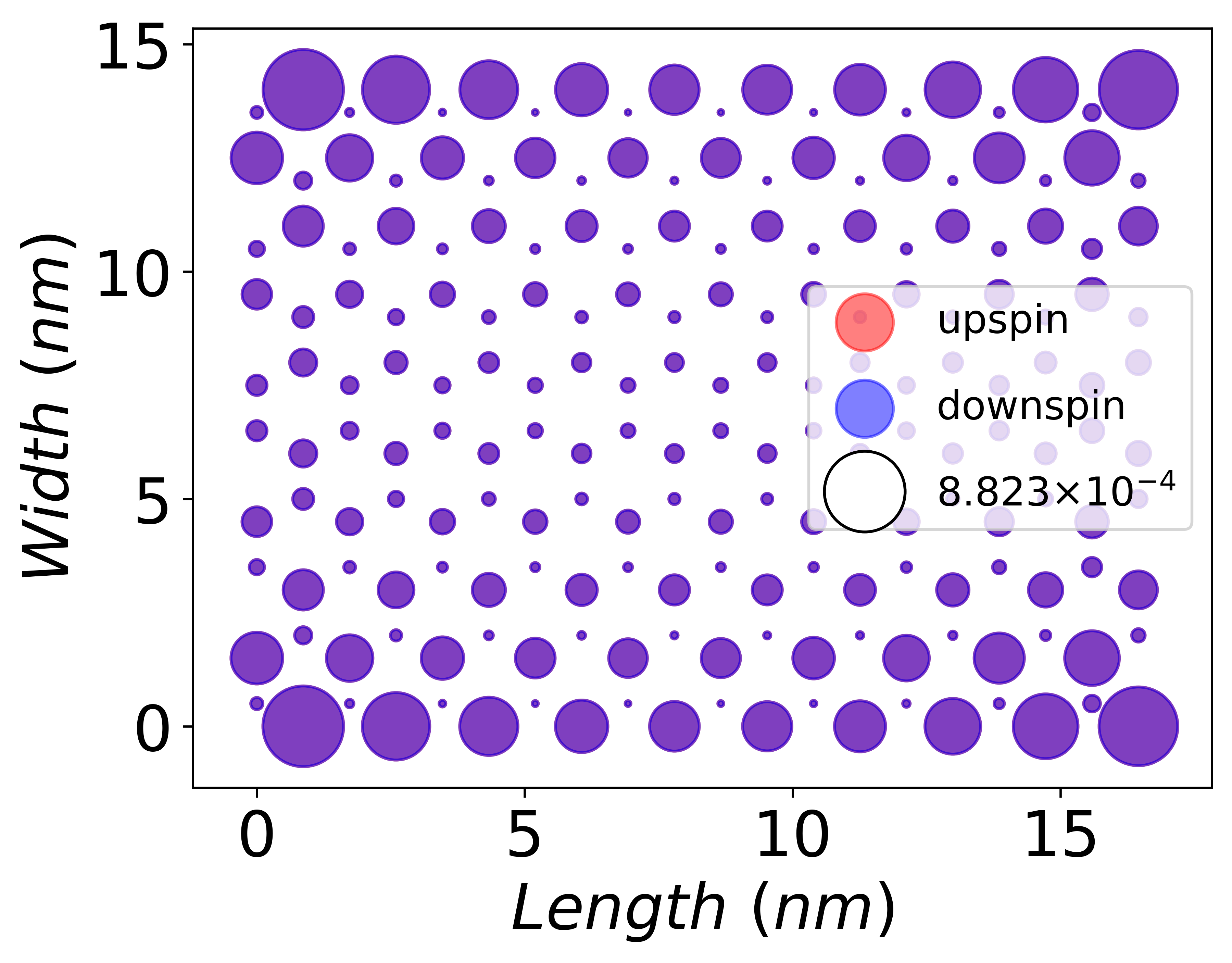}}
\hfill
\subfigure[]{\label{fig:ldos-qvh}
    \includegraphics[width=0.7\columnwidth]{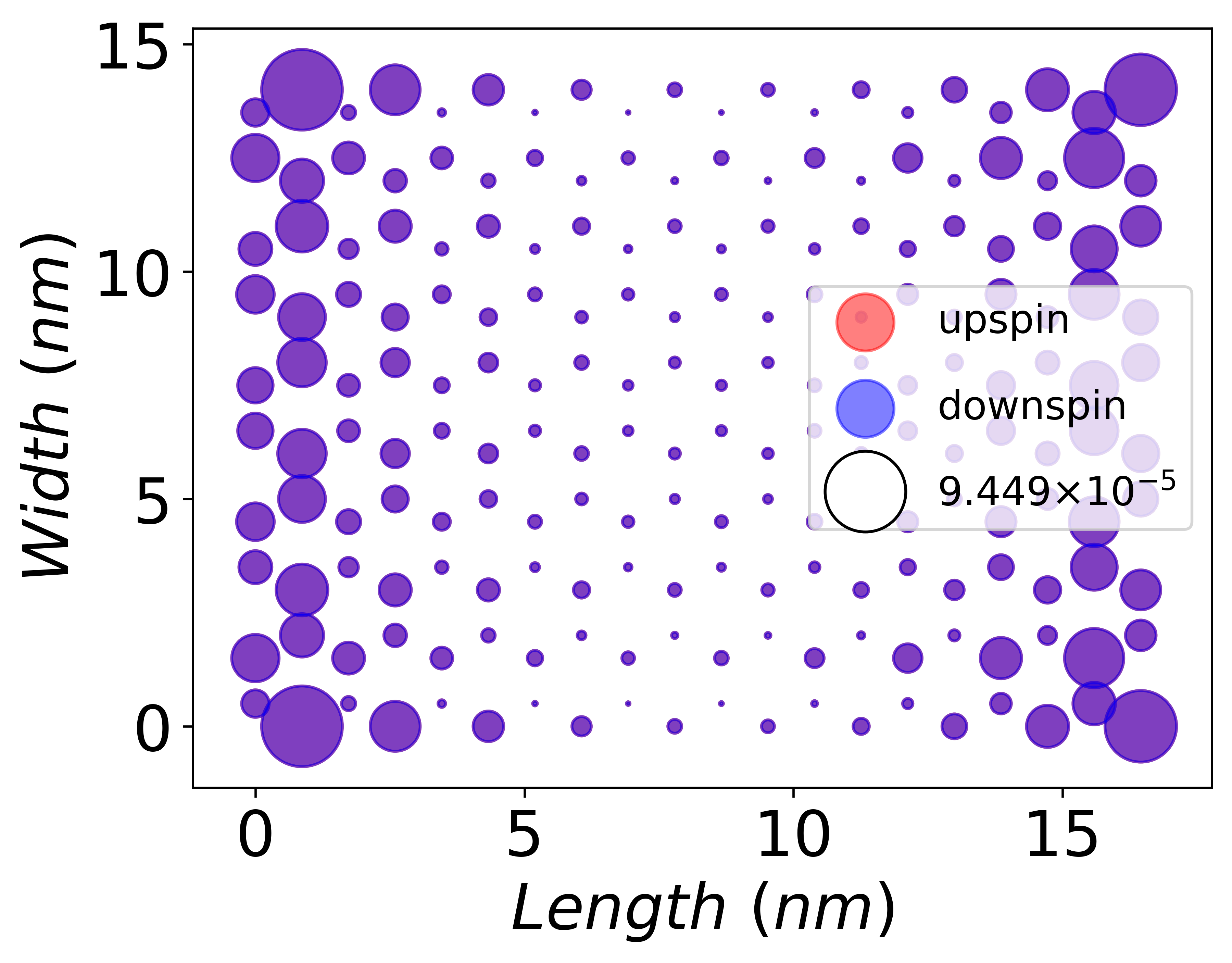}}
\caption{\label{fig:ldos}LDOS profiles of the (a) QSH, (b) SVPM, and (c) QVH phases of at the Fermi energy. The QSH phase has two conducting edge states of different spins, moving along opposite directions. The SVPM phase shows the onset of the phase transition of the current carrying edge states into bulk bound states. The QVH phase does not have any current carrying states and only has some states spilled over from the two contacts. }
\end{figure}
\begin{figure}
\includegraphics[width=0.8\columnwidth]{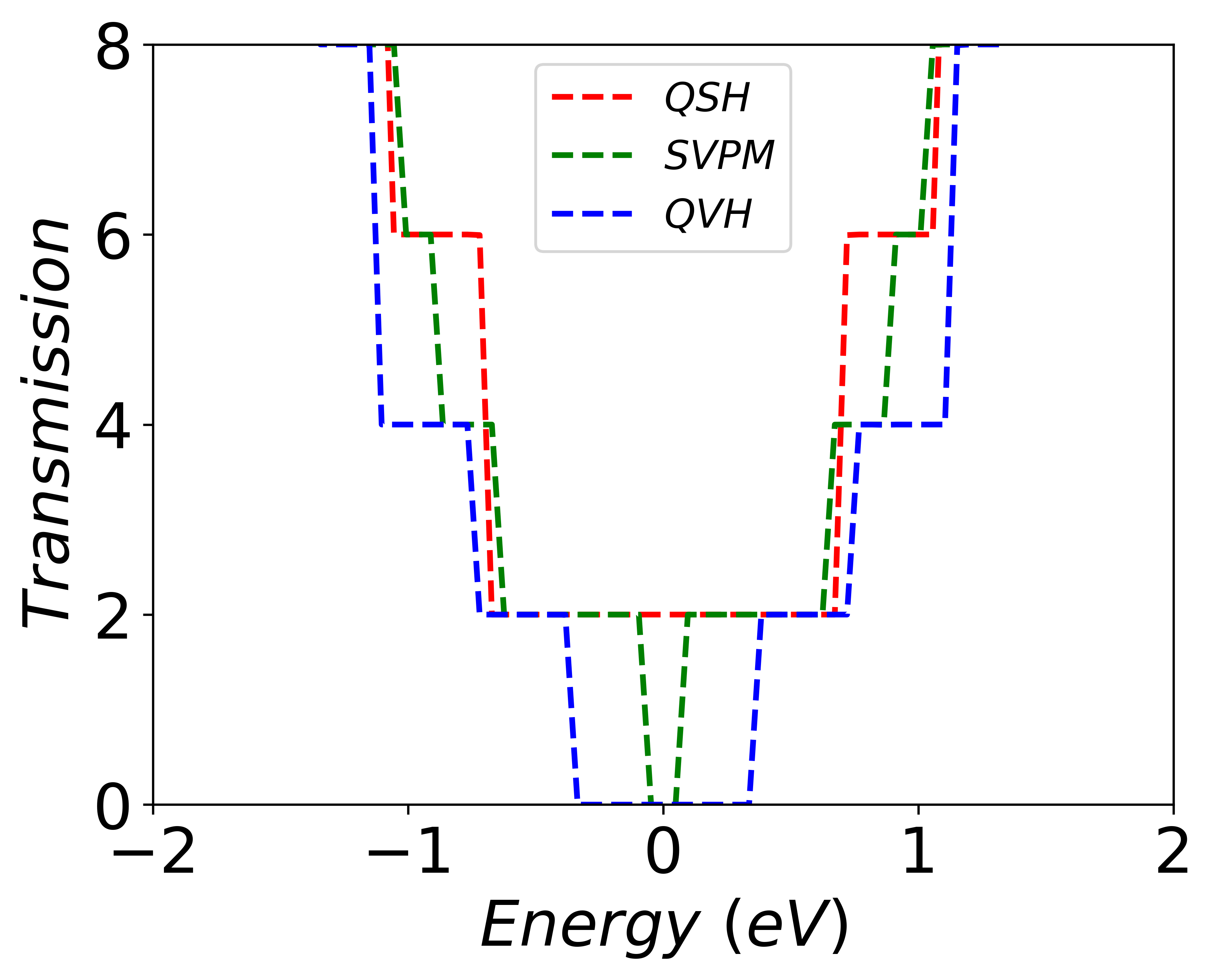}
\caption{\label{fig:trans} Transmission spectra in the coherent regime $d_m=0$ for the QSH, the SVPM, and the QVH phases. The two opposite spin edge states produce a conductance of $G_0$ at the Fermi energy for the QSH phase, whereas the absence of such states produces zero transmission at the Fermi energy for the QVH phase. The SVPM phase is where the transmission at Fermi energy falls to zero and a gap starts to open.}
\end{figure}
\begin{figure}
\includegraphics[width=0.8\columnwidth]{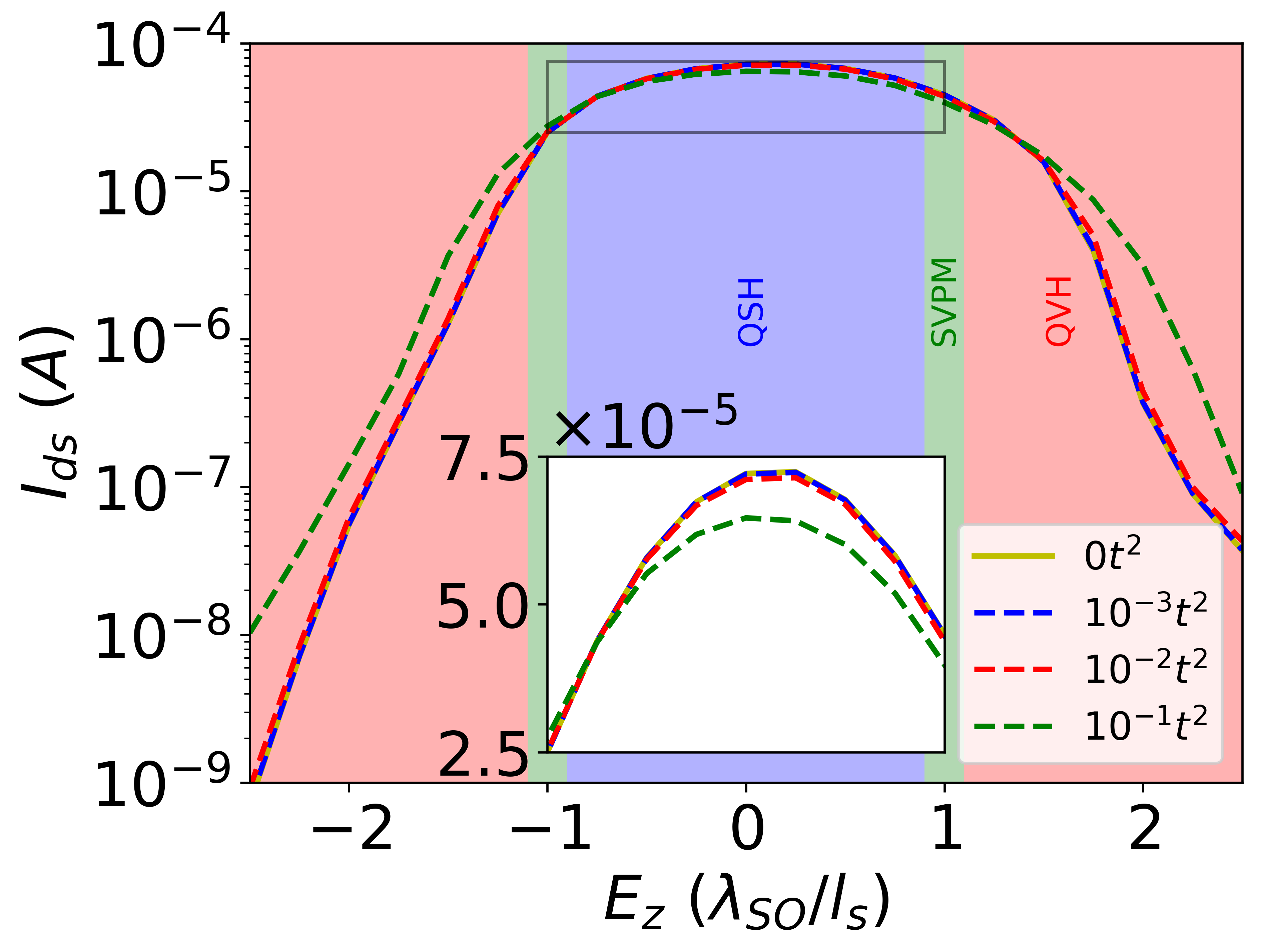}
\caption{\label{fig:Ids-vs-Ez} Current-field $I_{ds}$ vs $E_z$ characteristics of Topological FET with an applied bias $V_{ds} = 1V$ at $300K$ for different strengths of momentum dephasing in the log-scale. The inset shows the same in the linear scale only for the QSH phase. The changes in $E_z$ result in changes in the topological phases, which are indicated in the figure. The device shows a transistor-like behaviour where QSH and QVH represent the ON and OFF currents respectively.}
\end{figure}
\begin{figure}
\subfigure[]{\label{fig:coherent-trans}
    \includegraphics[width=0.6\columnwidth]{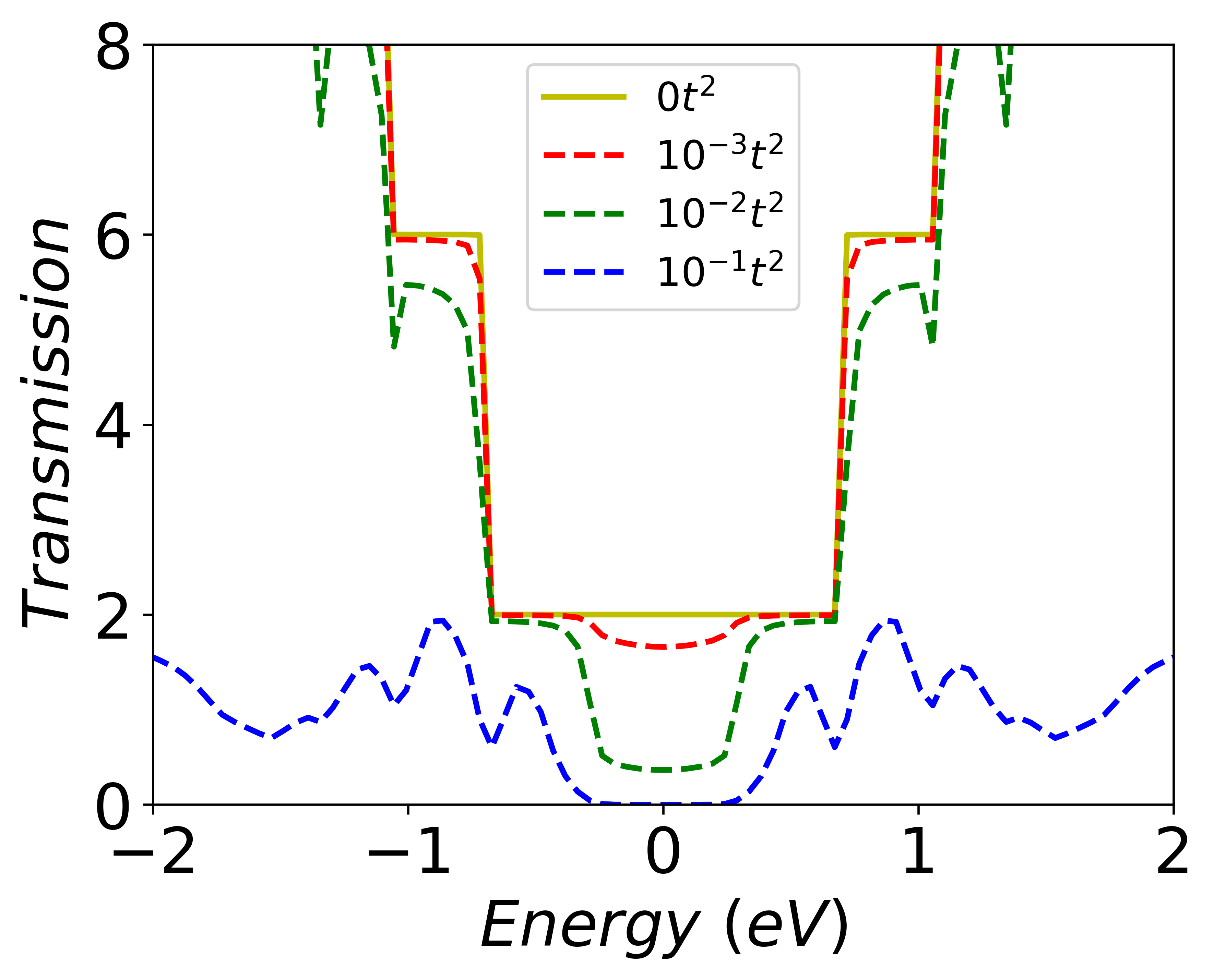}}
\hfill
\subfigure[]{\label{fig:effective-trans}
    \includegraphics[width=0.6\columnwidth]{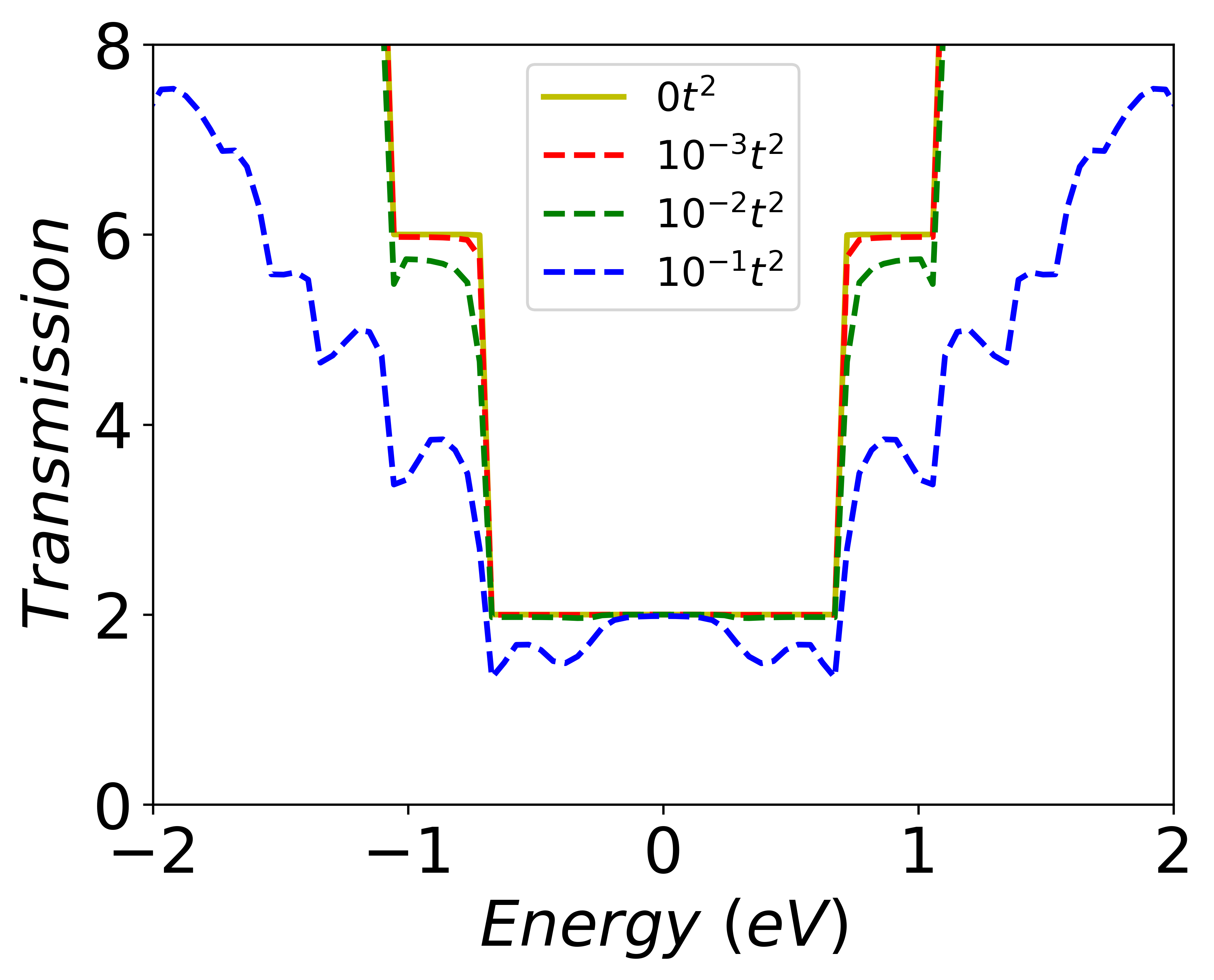}}
\hfill
\subfigure[]{\label{fig:coherent-ldos}
    \includegraphics[width=0.6\columnwidth]{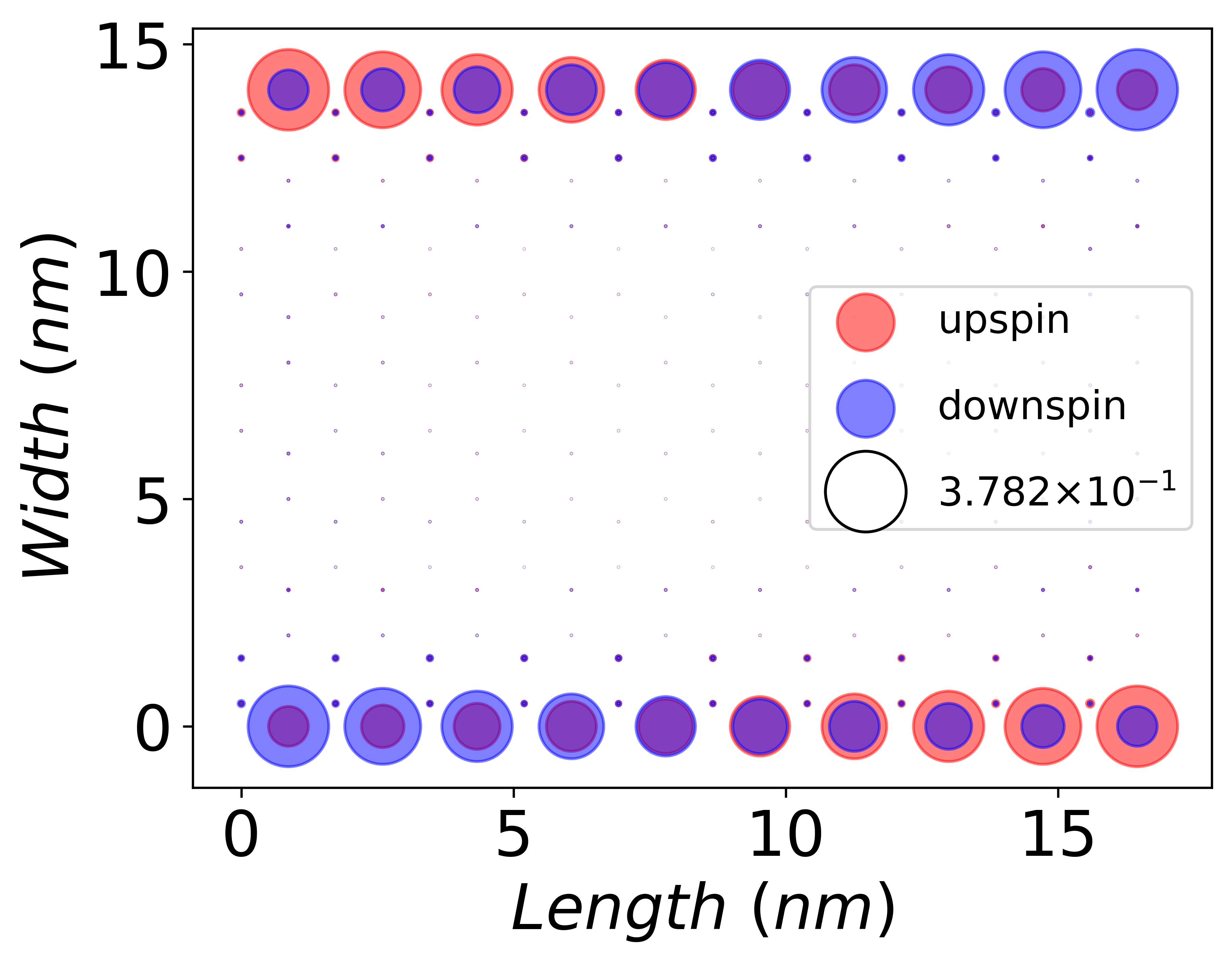}}
\caption{\label{fig:relaxation-effects} Transmission spectra under dephasing: (a) coherent transmission $T_{coh}$ and (b) the effective transmission $T_{eff}$ of the QSH phase with varying strengths of dephasing $d_m$. For identical dephasing strengths, the effective transmission $T_{eff}$ demonstrates more realistic profile in comparison with $T_{coh}$, indicating that the edge states are quite robust against dephasing. (c) LDOS of the QSH phase at Fermi energy with the momentum dephasing strength of $d_m = 10^{-2}t^2$. The decay of the different spin edge states demonstrates that the up-spin (down-spin) states are moving in the (counter) clockwise direction.}
\end{figure}
The local density of states (LDOS) in the device region of our structure is shown is Fig.~\ref{fig:ldos}. The left and right leads are connected to the central part through the self-energies $\Sigma^{r}_{L,R}$ defined via the NEGF formalism, and account for the quasi-continuum contribution of the electronic states from the left and right leads. The device Hamiltonian determines where these states will be present in the central part. The density of states (DOS) at the Fermi energy for the QSH phase reveals the presence of two edge-states along the edges of nanoribbon as shown in Fig.~\ref{fig:ldos-qsh}. Despite there being no bulk states in this phase, current carrying states are due to the presence of edge states. The edge states here consists of one up-spin (red) and one down-spin (blue) edge states on each of the two edges of the material. Interestingly, the up-spin states move clockwise along the edges and the down-spin states move counter-clockwise. Since, there is no relaxation or decay of states at this point they overlap perfectly and are seen here together in purple shade in Fig.~\ref{fig:ldos-qsh}. \\
\indent While transitioning from the QSH phase to the SVPM phase the two edge states are maintained with gradual decay in their LDOS magnitude. The SVPM phase, as shown in Fig.~\ref{fig:ldos-svpm}, is a boundary phase between the QSH and the QVH phase, where the edge states completely vanish. The QSH and the QVH states thus represent the ON and OFF states of the FET device respectively. In the QVH phase, at the Fermi energy, the only contribution to the very small LDOS comes from the two leads and is localised in the small area near both leads as shown in Fig.~\ref{fig:ldos-qvh}. In this phase, there are almost no current carrying states in the channel and the nanoribbon acts as a band insulator. The transmission characteristics of the device for different phases is shown in Fig.~\ref{fig:trans}. In the QSH phase (red-dash lines), there are two edge current carrying channels of opposite spins, with a total conductance $G_0$, where $G_0=2e^2/h$ is the conductance quantum. This remains with increasing electric fields, until the transition to the SVPM phase (green dashed lines), where the gap opens up. In the QVH phase, the device remains non-conductive for a range of energies above and below the Fermi energy (blue dash lines), due to the absence of any edge states.
\subsection{\label{sec:releff}Relaxation effects}
The dephasing model considered here is based on \eqref{eqn:momentum-dephasing=2}, which represents spatially uncorrelated scattering potentials. We can view this in momentum space as something that couples all momentum states and hence results in momentum relaxation. Due to this, we can also account for incoherent backscattering using this model. We use the current formula in \eqref{curr_operator} to evaluate the current drain current-electric field $I_{ds}-E_z$ characteristics of the device in Fig.~\ref{fig:Ids-vs-Ez}, when a source drain voltage of $V_{ds}= 1 V$ is applied. Here, $I_{ds}$ is the terminal current evaluated using \eqref{curr_operator} at either the left or the right contact renamed as the source and drain respectively, when a voltage is symmetrically applied such that $\mu_L= -\mu_R = qV_{ds}/2$. The plots in this section include the coherent limit and the non-coherent limit with varying amounts of momentum relaxation $d_m$ added. \\
\indent The $I_{ds}-E_z$ characteristics show that the device can be made to undergo phase transitions by applying the perpendicular electric field and turning it on or off by controlling the gate voltage, much like in a FET device. The current through the device in the ON-state QSH phase remains almost the same, only decreasing a bit with the advent of the SVPM phase and then decreasing rapidly in the OFF-state QVH phase. The device acts as a depletion mode FET, where the conductance falls with increasing magnitude of the gate voltage. There is an asymmetry in the current characteristics between the positive and negative gate bias. This is caused as a result of the non-zero buckling height of the 2D channel and the perpendicular electric field applied to the device through the gate. \\
\indent With the introduction of momentum relaxation by varying the parameter $d_m$, the edge states undergo scattering and the overall conductance is bound to decrease. This is seen in the $I_d-E_z$ plot, as the decreasing current for QSH region with increasing $d_m$. 
The coherent transmission $T_{coh}$ for different strengths of momentum relaxation, as shown in Fig.~\ref{fig:coherent-trans}, depicts that the edge states are indeed encountering scattering and the pristine states decay with increasing dephasing strengths. The LDOS plot with a dephasing strength of $d_m=10^{-2}t^2$ shows that the left going up-spin state and the right going down-spin state of the top edge and vice-versa for the bottom edge decays via edge scattering the further it goes into the channel as shown in Fig.~\ref{fig:coherent-ldos}. However, the effective transmission $T_{eff}$ for different dephasing strengths, as shown in Fig.~\ref{fig:effective-trans}, shows that even with dephasing present in the form of momentum relaxation, the pristine nature of the QSH current carrying states are more likely to be preserved and will conduct close to the ballistic limit with a conductance of $G_0$. This indicates the suppression of back-scattering in the edge states as these edge states are only allowed to conduct in a particular direction for a particular spin. And since, there is no nearby state with the same spin and opposite electron velocity where they can jump to, these edge states are topologically protected, up to a certain degree of dephasing. This aspect has been very well captured via the metric $T_{eff}$, that is rigorously derived from the current operator. It is also clear that merely using a transmission metric $T_{coh}$ inspired by the Landauer like form, is insufficient to capture the topological stability of the QSH phase. The effective transmission gives a more realistic picture of the actual current flow process as it is derived from the net current operator flowing into a terminal and encompasses all the scattering effects through in-scattering and electron correlation operators.\\
\begin{figure}
\includegraphics[width=0.8\columnwidth]{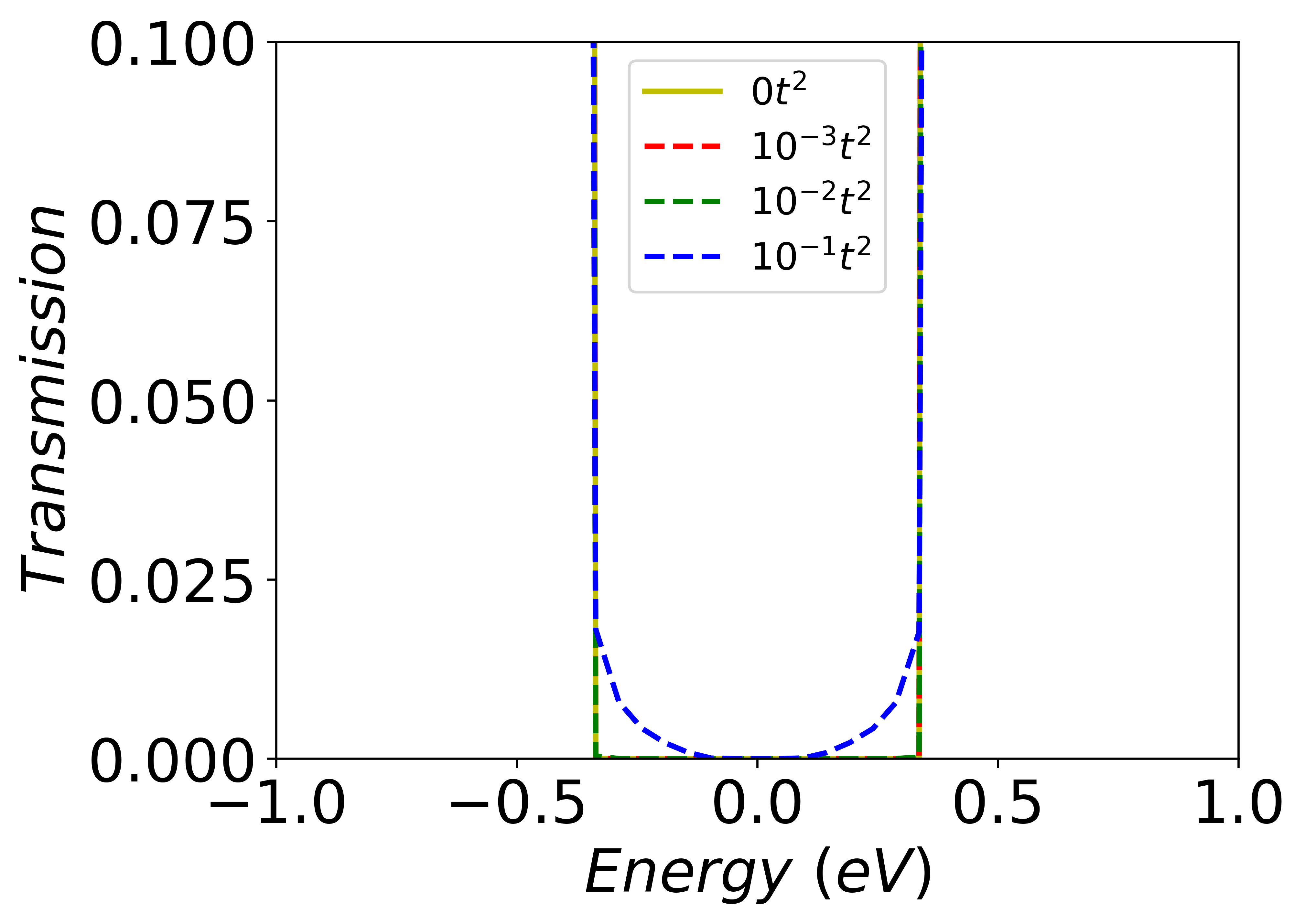}
\caption{\label{fig:teff-scattering-states-qvh} band tails: The effective transmission spectrum $T_{eff}$ of the QVH phase with varying dephasing strengths. We note that at larger values of dephasing, a distinct band tail effect is seen. Despite having a bandgap, the scattering states introduce a small amount of transmission within the gap, which in turn will affect the OFF current and hence reduce the ON-OFF ratio.}
\end{figure}
\indent We now consider the OFF-state described by the QVH phase. In any device, the ON-OFF ratio must be maintained high in order to ensure a reliable performance. Although, the QVH phase is supposed to be insulating due to the presence of a bandgap of $0.723eV$, a small current is observed in addition to the leakage current expected after the introduction of momentum dephasing, as seen in Fig~\ref{fig:Ids-vs-Ez}. The magnitude of this parasitic leakage current seems to increase with the increasing dephasing strength. Since the absence of almost any electronic states in the bandgap region, all method of current conduction is exhausted save for a very negligible amount of tunnelling. Looking back at the current operator \eqref{curr_operator}, we note that the total current can be composed as a sum of the regular ballistic current and the scattering current. The scattering current indeed includes a DOS contribution ($A_s(E) = {\bf{Tr}}\left([G^r][\Gamma_s][G^r] \right )$) and an electron correlation component ($G^{n}_s(E)={\bf{Tr}}\left(-i[G^r][\Sigma^{<}_s][G^r] \right )$) that are non-zero. These scatterer induced states \cite{Tillmann} can be connected to the lead energy levels, which are also above the same bandgap as the channel. Thus, we notice that a non-zero effective transmission is present in the QVH phase in presence of scattering relaxation, as shown in Fig~\ref{fig:teff-scattering-states-qvh}, which is well related to the band-tail formation, whose role is detrimental to the device performance \cite{Tillmann}. This transmission is larger close to the band-edges where the effect of such a level broadening is greater and falls off towards the middle of the gap.\\
\indent The rapid fall of the device current in QVH regime clearly indicates a sharp transition of the device from ON-state to OFF-state and hence points toward the feasibility of a good transistor device. Apart from the ON-OFF ratios analyzed here, it is also important to evaluate the sub-threshold slope characteristic of the transition. Although the sub-threshold slope is an important parameter for any transistor, it is difficult to analyze it without a detailed understanding of the gate electrostatics. As it depends on many parameters, they can be adjusted to control the sub-threshold slope. Topological switching of the kind analyzed in this paper has possible unique electrostatic signatures \cite{Nadeem-2021,Fuhrer} that may signal a smaller sub-threshold slope. We leave the detailed analysis of the subthreshold characteristics as well as the role of detailed device electrostatics and scattering for a detailed future work.
\section{\label{sec:conclusion}Conclusion}
This paper dealt with a detailed analysis of the electric field driven topological field effect transition between the QSH phase and the QVH phase in 2D-Xene based FET structures using the Keldysh non-equilibrium Green's function technique. Details of the transition with applied electric field were elucidated for the ON-OFF characteristics, with emphasis on the transport properties. A deeper analysis of the stability of the edge states as well as the nature of the switching process was considered in detail. We showed that, for moderate momentum relaxation, the current carrying QSH edge states are still pristine and show moderate decay with propagation. We also defined a metric, the effective transmission that is rigorously derived from the quantum mechanical current operator which paints a realistic picture when it comes to evaluating the stability of the topological edge states. Further, we also pointed out the important aspect of band-tail formation, which can lead to band-gap narrowing effects in the QVH phase, a remarkable effect that can arise even in the absence of in-elastic scattering effects that can lead to a degradation in the performance by enhancing parasitic OFF currents. The methods detailed in this paper can also be extended to include various other important effects like quantum dot formation and the related electron-electron interactions, which typically cause phase randomization without momentum relaxation, leading toward realistic modeling of topological field effect devices for various applications. 
\begin{acknowledgments}
The authors acknowledge Tillmann Kubis for insightful discussions. The research and development work undertaken in the project under the Visvesvaraya Ph.D Scheme of the Ministry of Electronics and Information Technology (MEITY), Government of India, is implemented by Digital India Corporation. This work is also supported by the Science and Engineering Research Board (SERB), Government of India, Grant No. STR/2019/000030, the Ministry of Human Resource Development (MHRD), Government of India, Grant No. STARS/APR2019/NS/226/FS under the STARS scheme.
\end{acknowledgments}
\section*{Data availability statement}
The data generated and/or analysed during the current study are not publicly available for legal/ethical reasons but are available from the corresponding author on reasonable request.
\bibliography{main}

\providecommand{\noopsort}[1]{}\providecommand{\singleletter}[1]{#1}%
\begin{thebibliography}{59}%
\makeatletter
\providecommand \@ifxundefined [1]{%
 \@ifx{#1\undefined}
}%
\providecommand \@ifnum [1]{%
 \ifnum #1\expandafter \@firstoftwo
 \else \expandafter \@secondoftwo
 \fi
}%
\providecommand \@ifx [1]{%
 \ifx #1\expandafter \@firstoftwo
 \else \expandafter \@secondoftwo
 \fi
}%
\providecommand \natexlab [1]{#1}%
\providecommand \enquote  [1]{``#1''}%
\providecommand \bibnamefont  [1]{#1}%
\providecommand \bibfnamefont [1]{#1}%
\providecommand \citenamefont [1]{#1}%
\providecommand \href@noop [0]{\@secondoftwo}%
\providecommand \href [0]{\begingroup \@sanitize@url \@href}%
\providecommand \@href[1]{\@@startlink{#1}\@@href}%
\providecommand \@@href[1]{\endgroup#1\@@endlink}%
\providecommand \@sanitize@url [0]{\catcode `\\12\catcode `\$12\catcode
  `\&12\catcode `\#12\catcode `\^12\catcode `\_12\catcode `\%12\relax}%
\providecommand \@@startlink[1]{}%
\providecommand \@@endlink[0]{}%
\providecommand \url  [0]{\begingroup\@sanitize@url \@url }%
\providecommand \@url [1]{\endgroup\@href {#1}{\urlprefix }}%
\providecommand \urlprefix  [0]{URL }%
\providecommand \Eprint [0]{\href }%
\providecommand \doibase [0]{https://doi.org/}%
\providecommand \selectlanguage [0]{\@gobble}%
\providecommand \bibinfo  [0]{\@secondoftwo}%
\providecommand \bibfield  [0]{\@secondoftwo}%
\providecommand \translation [1]{[#1]}%
\providecommand \BibitemOpen [0]{}%
\providecommand \bibitemStop [0]{}%
\providecommand \bibitemNoStop [0]{.\EOS\space}%
\providecommand \EOS [0]{\spacefactor3000\relax}%
\providecommand \BibitemShut  [1]{\csname bibitem#1\endcsname}%
\let\auto@bib@innerbib\@empty
\bibitem [{\citenamefont {Steinberg}\ \emph {et~al.}(2010)\citenamefont
  {Steinberg}, \citenamefont {Gardner}, \citenamefont {Lee},\ and\
  \citenamefont {Jarillo-Herrero}}]{Steinberg-2010}%
  \BibitemOpen
  \bibfield  {author} {\bibinfo {author} {\bibfnamefont {H.}~\bibnamefont
  {Steinberg}}, \bibinfo {author} {\bibfnamefont {D.~R.}\ \bibnamefont
  {Gardner}}, \bibinfo {author} {\bibfnamefont {Y.~S.}\ \bibnamefont {Lee}},\
  and\ \bibinfo {author} {\bibfnamefont {P.}~\bibnamefont {Jarillo-Herrero}},\
  }\href {https://doi.org/10.1021/nl1032183} {\bibfield  {journal} {\bibinfo
  {journal} {Nano Letters}\ }\textbf {\bibinfo {volume} {10}},\ \bibinfo
  {pages} {5032} (\bibinfo {year} {2010})}\BibitemShut {NoStop}%
\bibitem [{\citenamefont {Kara}\ \emph {et~al.}(2012)\citenamefont {Kara},
  \citenamefont {Enriquez}, \citenamefont {Seitsonen}, \citenamefont {{Lew Yan
  Voon}}, \citenamefont {Vizzini}, \citenamefont {Aufray},\ and\ \citenamefont
  {Oughaddou}}]{Sireview3}%
  \BibitemOpen
  \bibfield  {author} {\bibinfo {author} {\bibfnamefont {A.}~\bibnamefont
  {Kara}}, \bibinfo {author} {\bibfnamefont {H.}~\bibnamefont {Enriquez}},
  \bibinfo {author} {\bibfnamefont {A.~P.}\ \bibnamefont {Seitsonen}}, \bibinfo
  {author} {\bibfnamefont {L.}~\bibnamefont {{Lew Yan Voon}}}, \bibinfo
  {author} {\bibfnamefont {S.}~\bibnamefont {Vizzini}}, \bibinfo {author}
  {\bibfnamefont {B.}~\bibnamefont {Aufray}},\ and\ \bibinfo {author}
  {\bibfnamefont {H.}~\bibnamefont {Oughaddou}},\ }\href
  {https://doi.org/10.1016/j.surfrep.2011.10.001} {\bibfield  {journal}
  {\bibinfo  {journal} {Surface Science Reports}\ }\textbf {\bibinfo {volume}
  {67}},\ \bibinfo {pages} {1} (\bibinfo {year} {2012})}\BibitemShut {NoStop}%
\bibitem [{\citenamefont {Gilbert}(2021)}]{Gilbert-2021}%
  \BibitemOpen
  \bibfield  {author} {\bibinfo {author} {\bibfnamefont {M.~J.}\ \bibnamefont
  {Gilbert}},\ }\href {https://doi.org/10.1038/s42005-021-00569-5} {\bibfield
  {journal} {\bibinfo  {journal} {Communications Physics}\ }\textbf {\bibinfo
  {volume} {4}},\ \bibinfo {pages} {70} (\bibinfo {year} {2021})}\BibitemShut
  {NoStop}%
\bibitem [{\citenamefont {Liu}\ and\ \citenamefont {Ye}(2011)}]{Liu-2011}%
  \BibitemOpen
  \bibfield  {author} {\bibinfo {author} {\bibfnamefont {H.}~\bibnamefont
  {Liu}}\ and\ \bibinfo {author} {\bibfnamefont {P.~D.}\ \bibnamefont {Ye}},\
  }\href {https://doi.org/10.1063/1.3622306} {\bibfield  {journal} {\bibinfo
  {journal} {Applied Physics Letters}\ }\textbf {\bibinfo {volume} {99}},\
  \bibinfo {pages} {052108} (\bibinfo {year} {2011})}\BibitemShut {NoStop}%
\bibitem [{\citenamefont {Chang}\ \emph {et~al.}(2012)\citenamefont {Chang},
  \citenamefont {Register},\ and\ \citenamefont {Banerjee}}]{Chang-2012}%
  \BibitemOpen
  \bibfield  {author} {\bibinfo {author} {\bibfnamefont {J.}~\bibnamefont
  {Chang}}, \bibinfo {author} {\bibfnamefont {L.~F.}\ \bibnamefont
  {Register}},\ and\ \bibinfo {author} {\bibfnamefont {S.~K.}\ \bibnamefont
  {Banerjee}},\ }\href {https://doi.org/10.1063/1.4770324} {\bibfield
  {journal} {\bibinfo  {journal} {Journal of Applied Physics}\ }\textbf
  {\bibinfo {volume} {112}},\ \bibinfo {pages} {124511} (\bibinfo {year}
  {2012})}\BibitemShut {NoStop}%
\bibitem [{\citenamefont {Ionescu}\ and\ \citenamefont
  {Riel}(2011)}]{Ionescu-2011}%
  \BibitemOpen
  \bibfield  {author} {\bibinfo {author} {\bibfnamefont {A.~M.}\ \bibnamefont
  {Ionescu}}\ and\ \bibinfo {author} {\bibfnamefont {H.}~\bibnamefont {Riel}},\
  }\href {https://doi.org/10.1038/nature10679} {\bibfield  {journal} {\bibinfo
  {journal} {Nature}\ }\textbf {\bibinfo {volume} {479}},\ \bibinfo {pages}
  {329} (\bibinfo {year} {2011})}\BibitemShut {NoStop}%
\bibitem [{\citenamefont {Fu}\ \emph {et~al.}(2014)\citenamefont {Fu},
  \citenamefont {Gao},\ and\ \citenamefont {Yao}}]{Fu-2014}%
  \BibitemOpen
  \bibfield  {author} {\bibinfo {author} {\bibfnamefont {H.-H.}\ \bibnamefont
  {Fu}}, \bibinfo {author} {\bibfnamefont {J.-H.}\ \bibnamefont {Gao}},\ and\
  \bibinfo {author} {\bibfnamefont {K.-L.}\ \bibnamefont {Yao}},\ }\href
  {https://doi.org/10.1088/0957-4484/25/22/225201} {\bibfield  {journal}
  {\bibinfo  {journal} {Nanotechnology}\ }\textbf {\bibinfo {volume} {25}},\
  \bibinfo {pages} {225201} (\bibinfo {year} {2014})}\BibitemShut {NoStop}%
\bibitem [{\citenamefont {Akhavan}\ \emph {et~al.}(2014)\citenamefont
  {Akhavan}, \citenamefont {Jolley}, \citenamefont {Umana-Membreno},
  \citenamefont {Antoszewski},\ and\ \citenamefont {Faraone}}]{Akhavan-2014}%
  \BibitemOpen
  \bibfield  {author} {\bibinfo {author} {\bibfnamefont {N.~D.}\ \bibnamefont
  {Akhavan}}, \bibinfo {author} {\bibfnamefont {G.}~\bibnamefont {Jolley}},
  \bibinfo {author} {\bibfnamefont {G.~A.}\ \bibnamefont {Umana-Membreno}},
  \bibinfo {author} {\bibfnamefont {J.}~\bibnamefont {Antoszewski}},\ and\
  \bibinfo {author} {\bibfnamefont {L.}~\bibnamefont {Faraone}},\ }\href
  {https://doi.org/10.1063/1.4894152} {\bibfield  {journal} {\bibinfo
  {journal} {Journal of Applied Physics}\ }\textbf {\bibinfo {volume} {116}},\
  \bibinfo {pages} {084508} (\bibinfo {year} {2014})}\BibitemShut {NoStop}%
\bibitem [{\citenamefont {Sarkar}\ \emph {et~al.}(2015)\citenamefont {Sarkar},
  \citenamefont {Xie}, \citenamefont {Liu}, \citenamefont {Cao}, \citenamefont
  {Kang}, \citenamefont {Gong}, \citenamefont {Kraemer}, \citenamefont
  {Ajayan},\ and\ \citenamefont {Banerjee}}]{Sarkar-2015}%
  \BibitemOpen
  \bibfield  {author} {\bibinfo {author} {\bibfnamefont {D.}~\bibnamefont
  {Sarkar}}, \bibinfo {author} {\bibfnamefont {X.}~\bibnamefont {Xie}},
  \bibinfo {author} {\bibfnamefont {W.}~\bibnamefont {Liu}}, \bibinfo {author}
  {\bibfnamefont {W.}~\bibnamefont {Cao}}, \bibinfo {author} {\bibfnamefont
  {J.}~\bibnamefont {Kang}}, \bibinfo {author} {\bibfnamefont {Y.}~\bibnamefont
  {Gong}}, \bibinfo {author} {\bibfnamefont {S.}~\bibnamefont {Kraemer}},
  \bibinfo {author} {\bibfnamefont {P.~M.}\ \bibnamefont {Ajayan}},\ and\
  \bibinfo {author} {\bibfnamefont {K.}~\bibnamefont {Banerjee}},\ }\href
  {https://doi.org/10.1038/nature15387} {\bibfield  {journal} {\bibinfo
  {journal} {Nature}\ }\textbf {\bibinfo {volume} {526}},\ \bibinfo {pages}
  {91} (\bibinfo {year} {2015})}\BibitemShut {NoStop}%
\bibitem [{\citenamefont {Vandenberghe}\ and\ \citenamefont
  {Fischetti}(2017)}]{Vandenberghe-2017}%
  \BibitemOpen
  \bibfield  {author} {\bibinfo {author} {\bibfnamefont {W.~G.}\ \bibnamefont
  {Vandenberghe}}\ and\ \bibinfo {author} {\bibfnamefont {M.~V.}\ \bibnamefont
  {Fischetti}},\ }\href {https://doi.org/10.1038/ncomms14184} {\bibfield
  {journal} {\bibinfo  {journal} {Nature Communications}\ }\textbf {\bibinfo
  {volume} {8}},\ \bibinfo {pages} {14184} (\bibinfo {year}
  {2017})}\BibitemShut {NoStop}%
\bibitem [{\citenamefont {Simchi}\ \emph {et~al.}(2018)\citenamefont {Simchi},
  \citenamefont {Simchi}, \citenamefont {Fardmanesh},\ and\ \citenamefont
  {Peeters}}]{Simchi-2018}%
  \BibitemOpen
  \bibfield  {author} {\bibinfo {author} {\bibfnamefont {H.}~\bibnamefont
  {Simchi}}, \bibinfo {author} {\bibfnamefont {M.}~\bibnamefont {Simchi}},
  \bibinfo {author} {\bibfnamefont {M.}~\bibnamefont {Fardmanesh}},\ and\
  \bibinfo {author} {\bibfnamefont {F.~M.}\ \bibnamefont {Peeters}},\ }\href
  {https://doi.org/10.1088/1361-648x/aac050} {\bibfield  {journal} {\bibinfo
  {journal} {Journal of Physics: Condensed Matter}\ }\textbf {\bibinfo {volume}
  {30}},\ \bibinfo {pages} {235303} (\bibinfo {year} {2018})}\BibitemShut
  {NoStop}%
\bibitem [{\citenamefont {Xu}\ \emph {et~al.}(2019{\natexlab{a}})\citenamefont
  {Xu}, \citenamefont {Chen}, \citenamefont {Wang}, \citenamefont {Liu},\ and\
  \citenamefont {Ma}}]{Xu-2019}%
  \BibitemOpen
  \bibfield  {author} {\bibinfo {author} {\bibfnamefont {Y.}~\bibnamefont
  {Xu}}, \bibinfo {author} {\bibfnamefont {Y.-R.}\ \bibnamefont {Chen}},
  \bibinfo {author} {\bibfnamefont {J.}~\bibnamefont {Wang}}, \bibinfo {author}
  {\bibfnamefont {J.-F.}\ \bibnamefont {Liu}},\ and\ \bibinfo {author}
  {\bibfnamefont {Z.}~\bibnamefont {Ma}},\ }\href
  {https://doi.org/10.1103/PhysRevLett.123.206801} {\bibfield  {journal}
  {\bibinfo  {journal} {Phys. Rev. Lett.}\ }\textbf {\bibinfo {volume} {123}},\
  \bibinfo {pages} {206801} (\bibinfo {year} {2019}{\natexlab{a}})}\BibitemShut
  {NoStop}%
\bibitem [{\citenamefont {Yang}\ \emph {et~al.}(2020)\citenamefont {Yang},
  \citenamefont {Lü},\ and\ \citenamefont {Xie}}]{Yang-2020}%
  \BibitemOpen
  \bibfield  {author} {\bibinfo {author} {\bibfnamefont {J.-E.}\ \bibnamefont
  {Yang}}, \bibinfo {author} {\bibfnamefont {X.-L.}\ \bibnamefont {Lü}},\ and\
  \bibinfo {author} {\bibfnamefont {H.}~\bibnamefont {Xie}},\ }\href
  {https://doi.org/10.1088/1367-2630/abbbd2} {\bibfield  {journal} {\bibinfo
  {journal} {New Journal of Physics}\ }\textbf {\bibinfo {volume} {22}},\
  \bibinfo {pages} {103018} (\bibinfo {year} {2020})}\BibitemShut {NoStop}%
\bibitem [{\citenamefont {Nadeem}\ \emph {et~al.}(2021)\citenamefont {Nadeem},
  \citenamefont {Di~Bernardo}, \citenamefont {Wang}, \citenamefont {Fuhrer},\
  and\ \citenamefont {Culcer}}]{Nadeem-2021}%
  \BibitemOpen
  \bibfield  {author} {\bibinfo {author} {\bibfnamefont {M.}~\bibnamefont
  {Nadeem}}, \bibinfo {author} {\bibfnamefont {I.}~\bibnamefont {Di~Bernardo}},
  \bibinfo {author} {\bibfnamefont {X.}~\bibnamefont {Wang}}, \bibinfo {author}
  {\bibfnamefont {M.~S.}\ \bibnamefont {Fuhrer}},\ and\ \bibinfo {author}
  {\bibfnamefont {D.}~\bibnamefont {Culcer}},\ }\href
  {https://doi.org/10.1021/acs.nanolett.1c00378} {\bibfield  {journal}
  {\bibinfo  {journal} {Nano Letters}\ }\textbf {\bibinfo {volume} {21}},\
  \bibinfo {pages} {3155} (\bibinfo {year} {2021})}\BibitemShut {NoStop}%
\bibitem [{\citenamefont {Tao}\ \emph {et~al.}(2015)\citenamefont {Tao},
  \citenamefont {Cinquanta}, \citenamefont {Chiappe}, \citenamefont
  {Grazianetti}, \citenamefont {Fanciulli}, \citenamefont {Dubey},
  \citenamefont {Molle},\ and\ \citenamefont {Akinwande}}]{Tao2015_SiFET}%
  \BibitemOpen
  \bibfield  {author} {\bibinfo {author} {\bibfnamefont {L.}~\bibnamefont
  {Tao}}, \bibinfo {author} {\bibfnamefont {E.}~\bibnamefont {Cinquanta}},
  \bibinfo {author} {\bibfnamefont {D.}~\bibnamefont {Chiappe}}, \bibinfo
  {author} {\bibfnamefont {C.}~\bibnamefont {Grazianetti}}, \bibinfo {author}
  {\bibfnamefont {M.}~\bibnamefont {Fanciulli}}, \bibinfo {author}
  {\bibfnamefont {M.}~\bibnamefont {Dubey}}, \bibinfo {author} {\bibfnamefont
  {A.}~\bibnamefont {Molle}},\ and\ \bibinfo {author} {\bibfnamefont
  {D.}~\bibnamefont {Akinwande}},\ }\href
  {https://doi.org/10.1038/nnano.2014.325} {\bibfield  {journal} {\bibinfo
  {journal} {Nature Nanotechnology}\ }\textbf {\bibinfo {volume} {10}},\
  \bibinfo {pages} {227} (\bibinfo {year} {2015})}\BibitemShut {NoStop}%
\bibitem [{\citenamefont {Collins}\ \emph {et~al.}(2018)\citenamefont
  {Collins}, \citenamefont {Tadich}, \citenamefont {Wu}, \citenamefont {Gomes},
  \citenamefont {Rodrigues}, \citenamefont {Liu}, \citenamefont {Hellerstedt},
  \citenamefont {Ryu}, \citenamefont {Tang}, \citenamefont {Mo}, \citenamefont
  {Adam}, \citenamefont {Yang}, \citenamefont {Fuhrer},\ and\ \citenamefont
  {Edmonds}}]{Fuhrer}%
  \BibitemOpen
  \bibfield  {author} {\bibinfo {author} {\bibfnamefont {J.~L.}\ \bibnamefont
  {Collins}}, \bibinfo {author} {\bibfnamefont {A.}~\bibnamefont {Tadich}},
  \bibinfo {author} {\bibfnamefont {W.}~\bibnamefont {Wu}}, \bibinfo {author}
  {\bibfnamefont {L.~C.}\ \bibnamefont {Gomes}}, \bibinfo {author}
  {\bibfnamefont {J.~N.~B.}\ \bibnamefont {Rodrigues}}, \bibinfo {author}
  {\bibfnamefont {C.}~\bibnamefont {Liu}}, \bibinfo {author} {\bibfnamefont
  {J.}~\bibnamefont {Hellerstedt}}, \bibinfo {author} {\bibfnamefont
  {H.}~\bibnamefont {Ryu}}, \bibinfo {author} {\bibfnamefont {S.}~\bibnamefont
  {Tang}}, \bibinfo {author} {\bibfnamefont {S.-K.}\ \bibnamefont {Mo}},
  \bibinfo {author} {\bibfnamefont {S.}~\bibnamefont {Adam}}, \bibinfo {author}
  {\bibfnamefont {S.~A.}\ \bibnamefont {Yang}}, \bibinfo {author}
  {\bibfnamefont {M.~S.}\ \bibnamefont {Fuhrer}},\ and\ \bibinfo {author}
  {\bibfnamefont {M.~T.}\ \bibnamefont {Edmonds}},\ }\href
  {https://doi.org/10.1038/s41586-018-0788-5} {\bibfield  {journal} {\bibinfo
  {journal} {Nature}\ }\textbf {\bibinfo {volume} {564}},\ \bibinfo {pages}
  {390} (\bibinfo {year} {2018})}\BibitemShut {NoStop}%
\bibitem [{\citenamefont {Ezawa}(2012{\natexlab{a}})}]{Ezawa-2012}%
  \BibitemOpen
  \bibfield  {author} {\bibinfo {author} {\bibfnamefont {M.}~\bibnamefont
  {Ezawa}},\ }\href {https://doi.org/10.1103/PhysRevLett.109.055502} {\bibfield
   {journal} {\bibinfo  {journal} {Phys. Rev. Lett.}\ }\textbf {\bibinfo
  {volume} {109}},\ \bibinfo {pages} {055502} (\bibinfo {year}
  {2012}{\natexlab{a}})}\BibitemShut {NoStop}%
\bibitem [{\citenamefont {Zheng}\ \emph {et~al.}(2020)\citenamefont {Zheng},
  \citenamefont {Xiang}, \citenamefont {Li}, \citenamefont {Yuan},
  \citenamefont {Chi},\ and\ \citenamefont {Guo}}]{Zheng-2020}%
  \BibitemOpen
  \bibfield  {author} {\bibinfo {author} {\bibfnamefont {J.}~\bibnamefont
  {Zheng}}, \bibinfo {author} {\bibfnamefont {Y.}~\bibnamefont {Xiang}},
  \bibinfo {author} {\bibfnamefont {C.}~\bibnamefont {Li}}, \bibinfo {author}
  {\bibfnamefont {R.}~\bibnamefont {Yuan}}, \bibinfo {author} {\bibfnamefont
  {F.}~\bibnamefont {Chi}},\ and\ \bibinfo {author} {\bibfnamefont
  {Y.}~\bibnamefont {Guo}},\ }\href
  {https://doi.org/10.1103/PhysRevApplied.14.034027} {\bibfield  {journal}
  {\bibinfo  {journal} {Phys. Rev. Applied}\ }\textbf {\bibinfo {volume}
  {14}},\ \bibinfo {pages} {034027} (\bibinfo {year} {2020})}\BibitemShut
  {NoStop}%
\bibitem [{\citenamefont {Ishida}\ and\ \citenamefont
  {Liebsch}(2020)}]{Ishida-2020}%
  \BibitemOpen
  \bibfield  {author} {\bibinfo {author} {\bibfnamefont {H.}~\bibnamefont
  {Ishida}}\ and\ \bibinfo {author} {\bibfnamefont {A.}~\bibnamefont
  {Liebsch}},\ }\href {https://doi.org/10.1103/PhysRevResearch.2.023242}
  {\bibfield  {journal} {\bibinfo  {journal} {Phys. Rev. Research}\ }\textbf
  {\bibinfo {volume} {2}},\ \bibinfo {pages} {023242} (\bibinfo {year}
  {2020})}\BibitemShut {NoStop}%
\bibitem [{\citenamefont {Qian}\ \emph {et~al.}(2014)\citenamefont {Qian},
  \citenamefont {Liu}, \citenamefont {Fu},\ and\ \citenamefont
  {Li}}]{Qian-2014}%
  \BibitemOpen
  \bibfield  {author} {\bibinfo {author} {\bibfnamefont {X.}~\bibnamefont
  {Qian}}, \bibinfo {author} {\bibfnamefont {J.}~\bibnamefont {Liu}}, \bibinfo
  {author} {\bibfnamefont {L.}~\bibnamefont {Fu}},\ and\ \bibinfo {author}
  {\bibfnamefont {J.}~\bibnamefont {Li}},\ }\href
  {https://doi.org/10.1126/science.1256815} {\bibfield  {journal} {\bibinfo
  {journal} {Science}\ }\textbf {\bibinfo {volume} {346}},\ \bibinfo {pages}
  {1344} (\bibinfo {year} {2014})}\BibitemShut {NoStop}%
\bibitem [{\citenamefont {Hsieh}\ \emph {et~al.}(2012)\citenamefont {Hsieh},
  \citenamefont {Lin}, \citenamefont {Liu}, \citenamefont {Duan}, \citenamefont
  {Bansil},\ and\ \citenamefont {Fu}}]{Hsieh-2012}%
  \BibitemOpen
  \bibfield  {author} {\bibinfo {author} {\bibfnamefont {T.~H.}\ \bibnamefont
  {Hsieh}}, \bibinfo {author} {\bibfnamefont {H.}~\bibnamefont {Lin}}, \bibinfo
  {author} {\bibfnamefont {J.}~\bibnamefont {Liu}}, \bibinfo {author}
  {\bibfnamefont {W.}~\bibnamefont {Duan}}, \bibinfo {author} {\bibfnamefont
  {A.}~\bibnamefont {Bansil}},\ and\ \bibinfo {author} {\bibfnamefont
  {L.}~\bibnamefont {Fu}},\ }\href {https://doi.org/10.1038/ncomms1969}
  {\bibfield  {journal} {\bibinfo  {journal} {Nature Communications}\ }\textbf
  {\bibinfo {volume} {3}},\ \bibinfo {pages} {982} (\bibinfo {year}
  {2012})}\BibitemShut {NoStop}%
\bibitem [{\citenamefont {Ezawa}(2014)}]{Ezawa-2014}%
  \BibitemOpen
  \bibfield  {author} {\bibinfo {author} {\bibfnamefont {M.}~\bibnamefont
  {Ezawa}},\ }\href {https://doi.org/10.1088/1367-2630/16/6/065015} {\bibfield
  {journal} {\bibinfo  {journal} {New Journal of Physics}\ }\textbf {\bibinfo
  {volume} {16}},\ \bibinfo {pages} {065015} (\bibinfo {year}
  {2014})}\BibitemShut {NoStop}%
\bibitem [{\citenamefont {Sun}\ and\ \citenamefont {Singh}(2017)}]{Extra-2}%
  \BibitemOpen
  \bibfield  {author} {\bibinfo {author} {\bibfnamefont {J.}~\bibnamefont
  {Sun}}\ and\ \bibinfo {author} {\bibfnamefont {D.~J.}\ \bibnamefont
  {Singh}},\ }\href {https://doi.org/10.1063/1.4975819} {\bibfield  {journal}
  {\bibinfo  {journal} {Journal of Applied Physics}\ }\textbf {\bibinfo
  {volume} {121}},\ \bibinfo {pages} {064301} (\bibinfo {year}
  {2017})}\BibitemShut {NoStop}%
\bibitem [{\citenamefont {Zhu}\ \emph {et~al.}(2019)\citenamefont {Zhu},
  \citenamefont {Richter}, \citenamefont {Yu}, \citenamefont {Ye},
  \citenamefont {Zeng},\ and\ \citenamefont {Li}}]{Extra-3}%
  \BibitemOpen
  \bibfield  {author} {\bibinfo {author} {\bibfnamefont {H.}~\bibnamefont
  {Zhu}}, \bibinfo {author} {\bibfnamefont {C.~A.}\ \bibnamefont {Richter}},
  \bibinfo {author} {\bibfnamefont {S.}~\bibnamefont {Yu}}, \bibinfo {author}
  {\bibfnamefont {H.}~\bibnamefont {Ye}}, \bibinfo {author} {\bibfnamefont
  {M.}~\bibnamefont {Zeng}},\ and\ \bibinfo {author} {\bibfnamefont
  {Q.}~\bibnamefont {Li}},\ }\href {https://doi.org/10.1063/1.5111180}
  {\bibfield  {journal} {\bibinfo  {journal} {Applied Physics Letters}\
  }\textbf {\bibinfo {volume} {115}},\ \bibinfo {pages} {073107} (\bibinfo
  {year} {2019})}\BibitemShut {NoStop}%
\bibitem [{\citenamefont {Hu}\ \emph {et~al.}(2018)\citenamefont {Hu},
  \citenamefont {Zhang}, \citenamefont {Li},\ and\ \citenamefont
  {Wang}}]{Hu-2018}%
  \BibitemOpen
  \bibfield  {author} {\bibinfo {author} {\bibfnamefont {G.}~\bibnamefont
  {Hu}}, \bibinfo {author} {\bibfnamefont {Y.}~\bibnamefont {Zhang}}, \bibinfo
  {author} {\bibfnamefont {L.}~\bibnamefont {Li}},\ and\ \bibinfo {author}
  {\bibfnamefont {Z.~L.}\ \bibnamefont {Wang}},\ }\href
  {https://doi.org/10.1021/acsnano.7b07996} {\bibfield  {journal} {\bibinfo
  {journal} {ACS Nano}\ }\textbf {\bibinfo {volume} {12}},\ \bibinfo {pages}
  {779} (\bibinfo {year} {2018})}\BibitemShut {NoStop}%
\bibitem [{\citenamefont {Vali}\ \emph {et~al.}(2015)\citenamefont {Vali},
  \citenamefont {Dideban},\ and\ \citenamefont {Moezi}}]{Extra-6}%
  \BibitemOpen
  \bibfield  {author} {\bibinfo {author} {\bibfnamefont {M.}~\bibnamefont
  {Vali}}, \bibinfo {author} {\bibfnamefont {D.}~\bibnamefont {Dideban}},\ and\
  \bibinfo {author} {\bibfnamefont {N.}~\bibnamefont {Moezi}},\ }\href
  {https://doi.org/10.1016/j.physe.2015.02.011} {\bibfield  {journal} {\bibinfo
   {journal} {Physica E: Low-dimensional Systems and Nanostructures}\ }\textbf
  {\bibinfo {volume} {69}},\ \bibinfo {pages} {360} (\bibinfo {year}
  {2015})}\BibitemShut {NoStop}%
\bibitem [{\citenamefont {Liu}\ \emph {et~al.}(2015)\citenamefont {Liu},
  \citenamefont {Zhang}, \citenamefont {Abdalla}, \citenamefont {Fazzio},\ and\
  \citenamefont {Zunger}}]{Extra-7}%
  \BibitemOpen
  \bibfield  {author} {\bibinfo {author} {\bibfnamefont {Q.}~\bibnamefont
  {Liu}}, \bibinfo {author} {\bibfnamefont {X.}~\bibnamefont {Zhang}}, \bibinfo
  {author} {\bibfnamefont {L.~B.}\ \bibnamefont {Abdalla}}, \bibinfo {author}
  {\bibfnamefont {A.}~\bibnamefont {Fazzio}},\ and\ \bibinfo {author}
  {\bibfnamefont {A.}~\bibnamefont {Zunger}},\ }\href
  {https://doi.org/10.1021/nl5043769} {\bibfield  {journal} {\bibinfo
  {journal} {Nano Letters}\ }\textbf {\bibinfo {volume} {15}},\ \bibinfo
  {pages} {1222} (\bibinfo {year} {2015})}\BibitemShut {NoStop}%
\bibitem [{\citenamefont {Xu}\ \emph {et~al.}(2019{\natexlab{b}})\citenamefont
  {Xu}, \citenamefont {Chen}, \citenamefont {Wang}, \citenamefont {Liu},\ and\
  \citenamefont {Ma}}]{Extra-9}%
  \BibitemOpen
  \bibfield  {author} {\bibinfo {author} {\bibfnamefont {Y.}~\bibnamefont
  {Xu}}, \bibinfo {author} {\bibfnamefont {Y.-R.}\ \bibnamefont {Chen}},
  \bibinfo {author} {\bibfnamefont {J.}~\bibnamefont {Wang}}, \bibinfo {author}
  {\bibfnamefont {J.-F.}\ \bibnamefont {Liu}},\ and\ \bibinfo {author}
  {\bibfnamefont {Z.}~\bibnamefont {Ma}},\ }\href
  {https://doi.org/10.1103/PhysRevLett.123.206801} {\bibfield  {journal}
  {\bibinfo  {journal} {Phys. Rev. Lett.}\ }\textbf {\bibinfo {volume} {123}},\
  \bibinfo {pages} {206801} (\bibinfo {year} {2019}{\natexlab{b}})}\BibitemShut
  {NoStop}%
\bibitem [{\citenamefont {Fan}\ \emph {et~al.}(2016)\citenamefont {Fan},
  \citenamefont {Kou}, \citenamefont {Upadhyaya}, \citenamefont {Shao},
  \citenamefont {Pan}, \citenamefont {Lang}, \citenamefont {Che}, \citenamefont
  {Tang}, \citenamefont {Montazeri}, \citenamefont {Murata}, \citenamefont
  {Chang}, \citenamefont {Akyol}, \citenamefont {Yu}, \citenamefont {Nie},
  \citenamefont {Wong}, \citenamefont {Liu}, \citenamefont {Wang},
  \citenamefont {Tserkovnyak},\ and\ \citenamefont {Wang}}]{Extra-8}%
  \BibitemOpen
  \bibfield  {author} {\bibinfo {author} {\bibfnamefont {Y.}~\bibnamefont
  {Fan}}, \bibinfo {author} {\bibfnamefont {X.}~\bibnamefont {Kou}}, \bibinfo
  {author} {\bibfnamefont {P.}~\bibnamefont {Upadhyaya}}, \bibinfo {author}
  {\bibfnamefont {Q.}~\bibnamefont {Shao}}, \bibinfo {author} {\bibfnamefont
  {L.}~\bibnamefont {Pan}}, \bibinfo {author} {\bibfnamefont {M.}~\bibnamefont
  {Lang}}, \bibinfo {author} {\bibfnamefont {X.}~\bibnamefont {Che}}, \bibinfo
  {author} {\bibfnamefont {J.}~\bibnamefont {Tang}}, \bibinfo {author}
  {\bibfnamefont {M.}~\bibnamefont {Montazeri}}, \bibinfo {author}
  {\bibfnamefont {K.}~\bibnamefont {Murata}}, \bibinfo {author} {\bibfnamefont
  {L.-T.}\ \bibnamefont {Chang}}, \bibinfo {author} {\bibfnamefont
  {M.}~\bibnamefont {Akyol}}, \bibinfo {author} {\bibfnamefont
  {G.}~\bibnamefont {Yu}}, \bibinfo {author} {\bibfnamefont {T.}~\bibnamefont
  {Nie}}, \bibinfo {author} {\bibfnamefont {K.~L.}\ \bibnamefont {Wong}},
  \bibinfo {author} {\bibfnamefont {J.}~\bibnamefont {Liu}}, \bibinfo {author}
  {\bibfnamefont {Y.}~\bibnamefont {Wang}}, \bibinfo {author} {\bibfnamefont
  {Y.}~\bibnamefont {Tserkovnyak}},\ and\ \bibinfo {author} {\bibfnamefont
  {K.~L.}\ \bibnamefont {Wang}},\ }\href
  {https://doi.org/10.1038/nnano.2015.294} {\bibfield  {journal} {\bibinfo
  {journal} {Nature Nanotechnology}\ }\textbf {\bibinfo {volume} {11}},\
  \bibinfo {pages} {352} (\bibinfo {year} {2016})}\BibitemShut {NoStop}%
\bibitem [{\citenamefont {Hor}\ \emph {et~al.}(2009)\citenamefont {Hor},
  \citenamefont {Richardella}, \citenamefont {Roushan}, \citenamefont {Xia},
  \citenamefont {Checkelsky}, \citenamefont {Yazdani}, \citenamefont {Hasan},
  \citenamefont {Ong},\ and\ \citenamefont {Cava}}]{Hor-2009}%
  \BibitemOpen
  \bibfield  {author} {\bibinfo {author} {\bibfnamefont {Y.~S.}\ \bibnamefont
  {Hor}}, \bibinfo {author} {\bibfnamefont {A.}~\bibnamefont {Richardella}},
  \bibinfo {author} {\bibfnamefont {P.}~\bibnamefont {Roushan}}, \bibinfo
  {author} {\bibfnamefont {Y.}~\bibnamefont {Xia}}, \bibinfo {author}
  {\bibfnamefont {J.~G.}\ \bibnamefont {Checkelsky}}, \bibinfo {author}
  {\bibfnamefont {A.}~\bibnamefont {Yazdani}}, \bibinfo {author} {\bibfnamefont
  {M.~Z.}\ \bibnamefont {Hasan}}, \bibinfo {author} {\bibfnamefont {N.~P.}\
  \bibnamefont {Ong}},\ and\ \bibinfo {author} {\bibfnamefont {R.~J.}\
  \bibnamefont {Cava}},\ }\href {https://doi.org/10.1103/PhysRevB.79.195208}
  {\bibfield  {journal} {\bibinfo  {journal} {Phys. Rev. B}\ }\textbf {\bibinfo
  {volume} {79}},\ \bibinfo {pages} {195208} (\bibinfo {year}
  {2009})}\BibitemShut {NoStop}%
\bibitem [{\citenamefont {Cho}\ \emph {et~al.}(2011)\citenamefont {Cho},
  \citenamefont {Butch}, \citenamefont {Paglione},\ and\ \citenamefont
  {Fuhrer}}]{Cho-2011}%
  \BibitemOpen
  \bibfield  {author} {\bibinfo {author} {\bibfnamefont {S.}~\bibnamefont
  {Cho}}, \bibinfo {author} {\bibfnamefont {N.~P.}\ \bibnamefont {Butch}},
  \bibinfo {author} {\bibfnamefont {J.}~\bibnamefont {Paglione}},\ and\
  \bibinfo {author} {\bibfnamefont {M.~S.}\ \bibnamefont {Fuhrer}},\ }\href
  {https://doi.org/10.1021/nl200017f} {\bibfield  {journal} {\bibinfo
  {journal} {Nano Letters}\ }\textbf {\bibinfo {volume} {11}},\ \bibinfo
  {pages} {1925} (\bibinfo {year} {2011})}\BibitemShut {NoStop}%
\bibitem [{\citenamefont {Checkelsky}\ \emph {et~al.}(2011)\citenamefont
  {Checkelsky}, \citenamefont {Hor}, \citenamefont {Cava},\ and\ \citenamefont
  {Ong}}]{Checkelsky-2011}%
  \BibitemOpen
  \bibfield  {author} {\bibinfo {author} {\bibfnamefont {J.~G.}\ \bibnamefont
  {Checkelsky}}, \bibinfo {author} {\bibfnamefont {Y.~S.}\ \bibnamefont {Hor}},
  \bibinfo {author} {\bibfnamefont {R.~J.}\ \bibnamefont {Cava}},\ and\
  \bibinfo {author} {\bibfnamefont {N.~P.}\ \bibnamefont {Ong}},\ }\href
  {https://doi.org/10.1103/PhysRevLett.106.196801} {\bibfield  {journal}
  {\bibinfo  {journal} {Phys. Rev. Lett.}\ }\textbf {\bibinfo {volume} {106}},\
  \bibinfo {pages} {196801} (\bibinfo {year} {2011})}\BibitemShut {NoStop}%
\bibitem [{\citenamefont {Bansal}\ \emph {et~al.}(2012)\citenamefont {Bansal},
  \citenamefont {Kim}, \citenamefont {Brahlek}, \citenamefont {Edrey},\ and\
  \citenamefont {Oh}}]{Bansal-2012}%
  \BibitemOpen
  \bibfield  {author} {\bibinfo {author} {\bibfnamefont {N.}~\bibnamefont
  {Bansal}}, \bibinfo {author} {\bibfnamefont {Y.~S.}\ \bibnamefont {Kim}},
  \bibinfo {author} {\bibfnamefont {M.}~\bibnamefont {Brahlek}}, \bibinfo
  {author} {\bibfnamefont {E.}~\bibnamefont {Edrey}},\ and\ \bibinfo {author}
  {\bibfnamefont {S.}~\bibnamefont {Oh}},\ }\href
  {https://doi.org/10.1103/PhysRevLett.109.116804} {\bibfield  {journal}
  {\bibinfo  {journal} {Phys. Rev. Lett.}\ }\textbf {\bibinfo {volume} {109}},\
  \bibinfo {pages} {116804} (\bibinfo {year} {2012})}\BibitemShut {NoStop}%
\bibitem [{\citenamefont {Taskin}\ \emph {et~al.}(2012)\citenamefont {Taskin},
  \citenamefont {Sasaki}, \citenamefont {Segawa},\ and\ \citenamefont
  {Ando}}]{Taskin-2012}%
  \BibitemOpen
  \bibfield  {author} {\bibinfo {author} {\bibfnamefont {A.~A.}\ \bibnamefont
  {Taskin}}, \bibinfo {author} {\bibfnamefont {S.}~\bibnamefont {Sasaki}},
  \bibinfo {author} {\bibfnamefont {K.}~\bibnamefont {Segawa}},\ and\ \bibinfo
  {author} {\bibfnamefont {Y.}~\bibnamefont {Ando}},\ }\href
  {https://doi.org/10.1103/PhysRevLett.109.066803} {\bibfield  {journal}
  {\bibinfo  {journal} {Phys. Rev. Lett.}\ }\textbf {\bibinfo {volume} {109}},\
  \bibinfo {pages} {066803} (\bibinfo {year} {2012})}\BibitemShut {NoStop}%
\bibitem [{\citenamefont {Thalmeier}\ and\ \citenamefont
  {Akbari}(2020)}]{Thalmeier-2020}%
  \BibitemOpen
  \bibfield  {author} {\bibinfo {author} {\bibfnamefont {P.}~\bibnamefont
  {Thalmeier}}\ and\ \bibinfo {author} {\bibfnamefont {A.}~\bibnamefont
  {Akbari}},\ }\href {https://doi.org/10.1103/PhysRevResearch.2.033002}
  {\bibfield  {journal} {\bibinfo  {journal} {Phys. Rev. Research}\ }\textbf
  {\bibinfo {volume} {2}},\ \bibinfo {pages} {033002} (\bibinfo {year}
  {2020})}\BibitemShut {NoStop}%
\bibitem [{\citenamefont {Zhu}\ \emph {et~al.}(2013)\citenamefont {Zhu},
  \citenamefont {Richter}, \citenamefont {Zhao}, \citenamefont {Bonevich},
  \citenamefont {Kimes}, \citenamefont {Jang}, \citenamefont {Yuan},
  \citenamefont {Li}, \citenamefont {Arab}, \citenamefont {Kirillov},
  \citenamefont {Maslar}, \citenamefont {Ioannou},\ and\ \citenamefont
  {Li}}]{Zhu-2013}%
  \BibitemOpen
  \bibfield  {author} {\bibinfo {author} {\bibfnamefont {H.}~\bibnamefont
  {Zhu}}, \bibinfo {author} {\bibfnamefont {C.~A.}\ \bibnamefont {Richter}},
  \bibinfo {author} {\bibfnamefont {E.}~\bibnamefont {Zhao}}, \bibinfo {author}
  {\bibfnamefont {J.~E.}\ \bibnamefont {Bonevich}}, \bibinfo {author}
  {\bibfnamefont {W.~A.}\ \bibnamefont {Kimes}}, \bibinfo {author}
  {\bibfnamefont {H.-J.}\ \bibnamefont {Jang}}, \bibinfo {author}
  {\bibfnamefont {H.}~\bibnamefont {Yuan}}, \bibinfo {author} {\bibfnamefont
  {H.}~\bibnamefont {Li}}, \bibinfo {author} {\bibfnamefont {A.}~\bibnamefont
  {Arab}}, \bibinfo {author} {\bibfnamefont {O.}~\bibnamefont {Kirillov}},
  \bibinfo {author} {\bibfnamefont {J.~E.}\ \bibnamefont {Maslar}}, \bibinfo
  {author} {\bibfnamefont {D.~E.}\ \bibnamefont {Ioannou}},\ and\ \bibinfo
  {author} {\bibfnamefont {Q.}~\bibnamefont {Li}},\ }\href
  {https://doi.org/10.1038/srep01757} {\bibfield  {journal} {\bibinfo
  {journal} {Scientific Reports}\ }\textbf {\bibinfo {volume} {3}},\ \bibinfo
  {pages} {1757} (\bibinfo {year} {2013})}\BibitemShut {NoStop}%
\bibitem [{\citenamefont {Satake}\ \emph {et~al.}(2018)\citenamefont {Satake},
  \citenamefont {Shiogai}, \citenamefont {Fujiwara},\ and\ \citenamefont
  {Tsukazaki}}]{Satake-2018}%
  \BibitemOpen
  \bibfield  {author} {\bibinfo {author} {\bibfnamefont {Y.}~\bibnamefont
  {Satake}}, \bibinfo {author} {\bibfnamefont {J.}~\bibnamefont {Shiogai}},
  \bibinfo {author} {\bibfnamefont {K.}~\bibnamefont {Fujiwara}},\ and\
  \bibinfo {author} {\bibfnamefont {A.}~\bibnamefont {Tsukazaki}},\ }\href
  {https://doi.org/10.1103/PhysRevB.98.125415} {\bibfield  {journal} {\bibinfo
  {journal} {Phys. Rev. B}\ }\textbf {\bibinfo {volume} {98}},\ \bibinfo
  {pages} {125415} (\bibinfo {year} {2018})}\BibitemShut {NoStop}%
\bibitem [{\citenamefont {Jana}\ and\ \citenamefont
  {Muralidharan}(2021)}]{jana2021robust}%
  \BibitemOpen
  \bibfield  {author} {\bibinfo {author} {\bibfnamefont {K.}~\bibnamefont
  {Jana}}\ and\ \bibinfo {author} {\bibfnamefont {B.}~\bibnamefont
  {Muralidharan}},\ }\href@noop {} {\bibfield  {journal} {\bibinfo  {journal}
  {ArXiv: 2107.13318}\ } (\bibinfo {year} {2021})},\ \Eprint
  {https://arxiv.org/abs/2107.13318} {2107.13318} \BibitemShut {NoStop}%
\bibitem [{\citenamefont {Ezawa}(2012{\natexlab{b}})}]{Ezawa_njp_2012}%
  \BibitemOpen
  \bibfield  {author} {\bibinfo {author} {\bibfnamefont {M.}~\bibnamefont
  {Ezawa}},\ }\href {https://doi.org/10.1088/1367-2630/14/3/033003} {\bibfield
  {journal} {\bibinfo  {journal} {New Journal of Physics}\ }\textbf {\bibinfo
  {volume} {14}},\ \bibinfo {pages} {033003} (\bibinfo {year}
  {2012}{\natexlab{b}})}\BibitemShut {NoStop}%
\bibitem [{\citenamefont {Black-Schaffer}\ and\ \citenamefont
  {Balatsky}(2012)}]{Annica}%
  \BibitemOpen
  \bibfield  {author} {\bibinfo {author} {\bibfnamefont {A.~M.}\ \bibnamefont
  {Black-Schaffer}}\ and\ \bibinfo {author} {\bibfnamefont {A.~V.}\
  \bibnamefont {Balatsky}},\ }\href
  {https://doi.org/10.1103/PhysRevB.85.121103} {\bibfield  {journal} {\bibinfo
  {journal} {Phys. Rev. B}\ }\textbf {\bibinfo {volume} {85}},\ \bibinfo
  {pages} {121103} (\bibinfo {year} {2012})}\BibitemShut {NoStop}%
\bibitem [{\citenamefont {Groth}\ \emph {et~al.}(2014)\citenamefont {Groth},
  \citenamefont {Wimmer}, \citenamefont {Akhmerov},\ and\ \citenamefont
  {Waintal}}]{Kwant}%
  \BibitemOpen
  \bibfield  {author} {\bibinfo {author} {\bibfnamefont {C.~W.}\ \bibnamefont
  {Groth}}, \bibinfo {author} {\bibfnamefont {M.}~\bibnamefont {Wimmer}},
  \bibinfo {author} {\bibfnamefont {A.~R.}\ \bibnamefont {Akhmerov}},\ and\
  \bibinfo {author} {\bibfnamefont {X.}~\bibnamefont {Waintal}},\ }\href@noop
  {} {\bibfield  {journal} {\bibinfo  {journal} {New Journal of Physics}\
  }\textbf {\bibinfo {volume} {16}},\ \bibinfo {pages} {063065} (\bibinfo
  {year} {2014})}\BibitemShut {NoStop}%
\bibitem [{\citenamefont {Rischau}\ \emph {et~al.}(2016)\citenamefont
  {Rischau}, \citenamefont {Ubaldini}, \citenamefont {Giannini},\ and\
  \citenamefont {van~der Beek}}]{Puddles_2016}%
  \BibitemOpen
  \bibfield  {author} {\bibinfo {author} {\bibfnamefont {C.~W.}\ \bibnamefont
  {Rischau}}, \bibinfo {author} {\bibfnamefont {A.}~\bibnamefont {Ubaldini}},
  \bibinfo {author} {\bibfnamefont {E.}~\bibnamefont {Giannini}},\ and\
  \bibinfo {author} {\bibfnamefont {C.~J.}\ \bibnamefont {van~der Beek}},\
  }\href {https://doi.org/10.1088/1367-2630/18/7/073024} {\bibfield  {journal}
  {\bibinfo  {journal} {New Journal of Physics}\ }\textbf {\bibinfo {volume}
  {18}},\ \bibinfo {pages} {073024} (\bibinfo {year} {2016})}\BibitemShut
  {NoStop}%
\bibitem [{\citenamefont {Balram}\ \emph {et~al.}(2019)\citenamefont {Balram},
  \citenamefont {Flensberg}, \citenamefont {Paaske},\ and\ \citenamefont
  {Rudner}}]{Balram}%
  \BibitemOpen
  \bibfield  {author} {\bibinfo {author} {\bibfnamefont {A.~C.}\ \bibnamefont
  {Balram}}, \bibinfo {author} {\bibfnamefont {K.}~\bibnamefont {Flensberg}},
  \bibinfo {author} {\bibfnamefont {J.}~\bibnamefont {Paaske}},\ and\ \bibinfo
  {author} {\bibfnamefont {M.~S.}\ \bibnamefont {Rudner}},\ }\href
  {https://doi.org/10.1103/PhysRevLett.123.246803} {\bibfield  {journal}
  {\bibinfo  {journal} {Phys. Rev. Lett.}\ }\textbf {\bibinfo {volume} {123}},\
  \bibinfo {pages} {246803} (\bibinfo {year} {2019})}\BibitemShut {NoStop}%
\bibitem [{\citenamefont {Golizadeh-Mojarad}\ and\ \citenamefont
  {Datta}(2007)}]{Roksana}%
  \BibitemOpen
  \bibfield  {author} {\bibinfo {author} {\bibfnamefont {R.}~\bibnamefont
  {Golizadeh-Mojarad}}\ and\ \bibinfo {author} {\bibfnamefont {S.}~\bibnamefont
  {Datta}},\ }\href {https://doi.org/10.1103/PhysRevB.75.081301} {\bibfield
  {journal} {\bibinfo  {journal} {Phys. Rev. B}\ }\textbf {\bibinfo {volume}
  {75}},\ \bibinfo {pages} {081301} (\bibinfo {year} {2007})}\BibitemShut
  {NoStop}%
\bibitem [{\citenamefont {Datta}(1997)}]{Datta}%
  \BibitemOpen
  \bibfield  {author} {\bibinfo {author} {\bibfnamefont {S.}~\bibnamefont
  {Datta}},\ }\href@noop {} {\emph {\bibinfo {title} {{Electronic Transport in
  Mesoscopic Systems}}}}\ (\bibinfo  {publisher} {Cambridge University Press},\
  \bibinfo {year} {1997})\BibitemShut {NoStop}%
\bibitem [{\citenamefont {Danielewicz}(1984)}]{DANIELEWICZ1984239}%
  \BibitemOpen
  \bibfield  {author} {\bibinfo {author} {\bibfnamefont {P.}~\bibnamefont
  {Danielewicz}},\ }\href {https://doi.org/10.1016/0003-4916(84)90092-7}
  {\bibfield  {journal} {\bibinfo  {journal} {Annals of Physics}\ }\textbf
  {\bibinfo {volume} {152}},\ \bibinfo {pages} {239 } (\bibinfo {year}
  {1984})}\BibitemShut {NoStop}%
\bibitem [{\citenamefont {Camsari}\ \emph {et~al.}(2020)\citenamefont
  {Camsari}, \citenamefont {Chowdhury},\ and\ \citenamefont
  {Datta}}]{camsari2020nonequilibrium}%
  \BibitemOpen
  \bibfield  {author} {\bibinfo {author} {\bibfnamefont {K.~Y.}\ \bibnamefont
  {Camsari}}, \bibinfo {author} {\bibfnamefont {S.}~\bibnamefont {Chowdhury}},\
  and\ \bibinfo {author} {\bibfnamefont {S.}~\bibnamefont {Datta}},\
  }\href@noop {} {\bibinfo {title} {The non-equilibrium green function (negf)
  method}} (\bibinfo {year} {2020}),\ \Eprint
  {https://arxiv.org/abs/2008.01275} {arXiv:2008.01275 [cond-mat.mes-hall]}
  \BibitemShut {NoStop}%
\bibitem [{\citenamefont {Duse}\ \emph {et~al.}(2021)\citenamefont {Duse},
  \citenamefont {Sriram}, \citenamefont {Gharavi}, \citenamefont {Baugh},\ and\
  \citenamefont {Muralidharan}}]{Duse_2021}%
  \BibitemOpen
  \bibfield  {author} {\bibinfo {author} {\bibfnamefont {C.}~\bibnamefont
  {Duse}}, \bibinfo {author} {\bibfnamefont {P.}~\bibnamefont {Sriram}},
  \bibinfo {author} {\bibfnamefont {K.}~\bibnamefont {Gharavi}}, \bibinfo
  {author} {\bibfnamefont {J.}~\bibnamefont {Baugh}},\ and\ \bibinfo {author}
  {\bibfnamefont {B.}~\bibnamefont {Muralidharan}},\ }\href
  {https://doi.org/10.1088/1361-648x/ac0d16} {\bibfield  {journal} {\bibinfo
  {journal} {Journal of Physics: Condensed Matter}\ }\textbf {\bibinfo {volume}
  {33}},\ \bibinfo {pages} {365301} (\bibinfo {year} {2021})}\BibitemShut
  {NoStop}%
\bibitem [{\citenamefont {{Sharma}}\ \emph {et~al.}(2016)\citenamefont
  {{Sharma}}, \citenamefont {{Tulapurkar}},\ and\ \citenamefont
  {{Muralidharan}}}]{Abhishek_TED}%
  \BibitemOpen
  \bibfield  {author} {\bibinfo {author} {\bibfnamefont {A.}~\bibnamefont
  {{Sharma}}}, \bibinfo {author} {\bibfnamefont {A.}~\bibnamefont
  {{Tulapurkar}}},\ and\ \bibinfo {author} {\bibfnamefont {B.}~\bibnamefont
  {{Muralidharan}}},\ }\href {https://doi.org/10.1109/TED.2016.2606354}
  {\bibfield  {journal} {\bibinfo  {journal} {IEEE Transactions on Electron
  Devices}\ }\textbf {\bibinfo {volume} {63}},\ \bibinfo {pages} {4527}
  (\bibinfo {year} {2016})}\BibitemShut {NoStop}%
\bibitem [{\citenamefont {Sharma}\ \emph {et~al.}(2017)\citenamefont {Sharma},
  \citenamefont {Tulapurkar},\ and\ \citenamefont
  {Muralidharan}}]{PhysRevApplied.8.064014}%
  \BibitemOpen
  \bibfield  {author} {\bibinfo {author} {\bibfnamefont {A.}~\bibnamefont
  {Sharma}}, \bibinfo {author} {\bibfnamefont {A.~A.}\ \bibnamefont
  {Tulapurkar}},\ and\ \bibinfo {author} {\bibfnamefont {B.}~\bibnamefont
  {Muralidharan}},\ }\href {https://doi.org/10.1103/PhysRevApplied.8.064014}
  {\bibfield  {journal} {\bibinfo  {journal} {Phys. Rev. Applied}\ }\textbf
  {\bibinfo {volume} {8}},\ \bibinfo {pages} {064014} (\bibinfo {year}
  {2017})}\BibitemShut {NoStop}%
\bibitem [{\citenamefont {Sharma}\ \emph {et~al.}(2018)\citenamefont {Sharma},
  \citenamefont {Tulapurkar},\ and\ \citenamefont
  {Muralidharan}}]{Abhishek_APL}%
  \BibitemOpen
  \bibfield  {author} {\bibinfo {author} {\bibfnamefont {A.}~\bibnamefont
  {Sharma}}, \bibinfo {author} {\bibfnamefont {A.~A.}\ \bibnamefont
  {Tulapurkar}},\ and\ \bibinfo {author} {\bibfnamefont {B.}~\bibnamefont
  {Muralidharan}},\ }\href {https://doi.org/10.1063/1.5023159} {\bibfield
  {journal} {\bibinfo  {journal} {Applied Physics Letters}\ }\textbf {\bibinfo
  {volume} {112}},\ \bibinfo {pages} {192404} (\bibinfo {year}
  {2018})}\BibitemShut {NoStop}%
\bibitem [{\citenamefont {Singha}\ and\ \citenamefont
  {Muralidharan}(2018)}]{Aniket_JAP}%
  \BibitemOpen
  \bibfield  {author} {\bibinfo {author} {\bibfnamefont {A.}~\bibnamefont
  {Singha}}\ and\ \bibinfo {author} {\bibfnamefont {B.}~\bibnamefont
  {Muralidharan}},\ }\href {https://doi.org/10.1063/1.5044254} {\bibfield
  {journal} {\bibinfo  {journal} {Journal of Applied Physics}\ }\textbf
  {\bibinfo {volume} {124}},\ \bibinfo {pages} {144901} (\bibinfo {year}
  {2018})}\BibitemShut {NoStop}%
\bibitem [{\citenamefont {Lahiri}\ \emph {et~al.}(2018)\citenamefont {Lahiri},
  \citenamefont {Gharavi}, \citenamefont {Baugh},\ and\ \citenamefont
  {Muralidharan}}]{Aritra}%
  \BibitemOpen
  \bibfield  {author} {\bibinfo {author} {\bibfnamefont {A.}~\bibnamefont
  {Lahiri}}, \bibinfo {author} {\bibfnamefont {K.}~\bibnamefont {Gharavi}},
  \bibinfo {author} {\bibfnamefont {J.}~\bibnamefont {Baugh}},\ and\ \bibinfo
  {author} {\bibfnamefont {B.}~\bibnamefont {Muralidharan}},\ }\href
  {https://doi.org/10.1103/PhysRevB.98.125417} {\bibfield  {journal} {\bibinfo
  {journal} {Phys. Rev. B}\ }\textbf {\bibinfo {volume} {98}},\ \bibinfo
  {pages} {125417} (\bibinfo {year} {2018})}\BibitemShut {NoStop}%
\bibitem [{\citenamefont {Sriram}\ \emph {et~al.}(2019)\citenamefont {Sriram},
  \citenamefont {Kalantre}, \citenamefont {Gharavi}, \citenamefont {Baugh},\
  and\ \citenamefont {Muralidharan}}]{Praveen}%
  \BibitemOpen
  \bibfield  {author} {\bibinfo {author} {\bibfnamefont {P.}~\bibnamefont
  {Sriram}}, \bibinfo {author} {\bibfnamefont {S.~S.}\ \bibnamefont
  {Kalantre}}, \bibinfo {author} {\bibfnamefont {K.}~\bibnamefont {Gharavi}},
  \bibinfo {author} {\bibfnamefont {J.}~\bibnamefont {Baugh}},\ and\ \bibinfo
  {author} {\bibfnamefont {B.}~\bibnamefont {Muralidharan}},\ }\href
  {https://doi.org/10.1103/PhysRevB.100.155431} {\bibfield  {journal} {\bibinfo
   {journal} {Phys. Rev. B}\ }\textbf {\bibinfo {volume} {100}},\ \bibinfo
  {pages} {155431} (\bibinfo {year} {2019})}\BibitemShut {NoStop}%
\bibitem [{\citenamefont {Meir}\ and\ \citenamefont
  {Wingreen}(1992)}]{Meir-Wingreen-1992}%
  \BibitemOpen
  \bibfield  {author} {\bibinfo {author} {\bibfnamefont {Y.}~\bibnamefont
  {Meir}}\ and\ \bibinfo {author} {\bibfnamefont {N.~S.}\ \bibnamefont
  {Wingreen}},\ }\href {https://doi.org/10.1103/PhysRevLett.68.2512} {\bibfield
   {journal} {\bibinfo  {journal} {Phys. Rev. Lett.}\ }\textbf {\bibinfo
  {volume} {68}},\ \bibinfo {pages} {2512} (\bibinfo {year}
  {1992})}\BibitemShut {NoStop}%
\bibitem [{\citenamefont {Overhauser}(1989)}]{Overhauser}%
  \BibitemOpen
  \bibfield  {author} {\bibinfo {author} {\bibfnamefont {A.~W.}\ \bibnamefont
  {Overhauser}},\ }\href@noop {} {\bibfield  {journal} {\bibinfo  {journal}
  {Applied Physics Letters}\ }\textbf {\bibinfo {volume} {54}},\ \bibinfo
  {pages} {2490} (\bibinfo {year} {1989})}\BibitemShut {NoStop}%
\bibitem [{\citenamefont {Sarangapani}\ \emph {et~al.}(2019)\citenamefont
  {Sarangapani}, \citenamefont {Chu}, \citenamefont {Charles}, \citenamefont
  {Klimeck},\ and\ \citenamefont {Kubis}}]{Tillmann}%
  \BibitemOpen
  \bibfield  {author} {\bibinfo {author} {\bibfnamefont {P.}~\bibnamefont
  {Sarangapani}}, \bibinfo {author} {\bibfnamefont {Y.}~\bibnamefont {Chu}},
  \bibinfo {author} {\bibfnamefont {J.}~\bibnamefont {Charles}}, \bibinfo
  {author} {\bibfnamefont {G.}~\bibnamefont {Klimeck}},\ and\ \bibinfo {author}
  {\bibfnamefont {T.}~\bibnamefont {Kubis}},\ }\href
  {https://doi.org/10.1103/PhysRevApplied.12.044045} {\bibfield  {journal}
  {\bibinfo  {journal} {Phys. Rev. Applied}\ }\textbf {\bibinfo {volume}
  {12}},\ \bibinfo {pages} {044045} (\bibinfo {year} {2019})}\BibitemShut
  {NoStop}%
\bibitem [{\citenamefont {Muralidharan}\ \emph {et~al.}(2008)\citenamefont
  {Muralidharan}, \citenamefont {Siddiqui},\ and\ \citenamefont
  {Ghosh}}]{Muralidharan_2008}%
  \BibitemOpen
  \bibfield  {author} {\bibinfo {author} {\bibfnamefont {B.}~\bibnamefont
  {Muralidharan}}, \bibinfo {author} {\bibfnamefont {L.}~\bibnamefont
  {Siddiqui}},\ and\ \bibinfo {author} {\bibfnamefont {A.~W.}\ \bibnamefont
  {Ghosh}},\ }\href {https://doi.org/10.1088/0953-8984/20/37/374109} {\bibfield
   {journal} {\bibinfo  {journal} {Journal of Physics: Condensed Matter}\
  }\textbf {\bibinfo {volume} {20}},\ \bibinfo {pages} {374109} (\bibinfo
  {year} {2008})}\BibitemShut {NoStop}%
\bibitem [{\citenamefont {Konschuh}\ \emph {et~al.}(2010)\citenamefont
  {Konschuh}, \citenamefont {Gmitra},\ and\ \citenamefont
  {Fabian}}]{Konschuh-2010}%
  \BibitemOpen
  \bibfield  {author} {\bibinfo {author} {\bibfnamefont {S.}~\bibnamefont
  {Konschuh}}, \bibinfo {author} {\bibfnamefont {M.}~\bibnamefont {Gmitra}},\
  and\ \bibinfo {author} {\bibfnamefont {J.}~\bibnamefont {Fabian}},\ }\href
  {https://doi.org/10.1103/PhysRevB.82.245412} {\bibfield  {journal} {\bibinfo
  {journal} {Phys. Rev. B}\ }\textbf {\bibinfo {volume} {82}},\ \bibinfo
  {pages} {245412} (\bibinfo {year} {2010})}\BibitemShut {NoStop}%
\end{thebibliography}%

\end{document}